\documentclass[a4paper,11pt]{article}

\usepackage{graphicx}
\graphicspath{{./}{figures/}}

\usepackage[square,numbers]{natbib}

\usepackage{authblk}

\begin{document}

\title{Intensities of the hydrogen Balmer lines of solar-like stars revealed by the LAMOST spectroscopic surveys}

\author[1,2]{Han He} 
\affil[1]{National Astronomical Observatories, Chinese Academy of Sciences, Beijing, 100101, China\ \ email: hehan@nao.cas.cn}
\affil[2]{University of Chinese Academy of Sciences, Beijing, 100049, China}

\date{}

\maketitle

\abstract
{The intensities of the hydrogen Balmer lines of solar-like stars are investigated for stellar chromospheric activity by using the co-source spectral data of the LAMOST Low-Resolution Spectroscopic Survey (LRS) and Medium-Resolution Spectroscopic Survey (MRS).
The Balmer H$\alpha$, H$\beta$, H$\gamma$, and H$\delta$ lines in the LRS data and the H$\alpha$ line in the MRS data are analyzed.
The absolute flux indexes, 
defined as the ratios of the absolute fluxes at the centers of the Balmer lines to the stellar bolometric flux, 
are employed to indicate the intensity magnitudes of the Balmer lines in response to stellar activity.
The H$\alpha$ indexes derived from the LRS data and the MRS data, respectively, are calibrated to be quantitatively consistent with each other.
It is found that, as the H$\alpha$ index increases, 
the H$\beta$, H$\gamma$, and H$\delta$ indexes first present trend of increasing and then decreasing, and finally increase synchronously with the H$\alpha$ index.
The distributions of the Balmer line indexes also reveal the three distinct stages of stellar activity (normal stage, intense stage, and extremely intense stage),
in which the extremely intense stage is characterized by the synchronous growth of the indexes of the four Balmer lines.
The different behaviors of the H$\beta$, H$\gamma$, and H$\delta$ lines from that of the H$\alpha$ line can be interpreted by the different mechanisms by which the line-core intensities are formed,
and the three distinct activity stages imply the very different magnetic field environments and physical conditions of solar-like stars. \\
\\
Keywords: stars: activity --- stars: chromospheres --- stars: solar-type}

\section{Introduction} \label{sec:intro}

The hydrogen is the most abundant element in solar-like stars \citep{2009ARA&A..47..481A}. 
The spectral lines in the Balmer series of hydrogen are among the most commonly used spectral features in the optical band for diagnosing the physical properties of astronomical objects. 
The intensities at the centers of Balmer lines are known to be sensitive to stellar magnetic activity \citep{2008LRSP....5....2H, 2017ARA&A..55..159L},
and the Balmer series is often used to analyze the characteristics of stellar chromospheres for both the steady state and flare activity \citep{1980ApJS...42..351D, 1991ApJ...378..725H, 2010AJ....140.1402H}. 
The Balmer decrement (relative strengths between the Balmer lines) is often utilized to infer the physical environment and plasma parameters in the source regions of stellar magnetic activity in the stellar chromosphere, 
especially during stellar flares \citep{1980ApJS...42..351D, 1991MmSAI..62..243B}. 

In order to investigate diagnostics of the Balmer lines for stellar chromospheric activity, Cram and Mullan \citep{1979ApJ...234..579C} computed a grid of model chromosphere for M dwarf stars, 
and calculated the spectral profiles for three Balmer lines: H$\alpha$, H$\beta$, and H$\gamma$.
Their modeling results showed that, when the chromosphere is absent on stars, these Balmer lines manifest as weak absorption lines; 
as the materials of chromosphere increase, the Balmer lines first become deeper,
then fill in to become shallower, and finally become emission lines. 
In a follow-up research, Cram and Mullan \citep{1985ApJ...294..626C} modeled the formation of the H$\alpha$ absorption line in chromospheres of cool stars, 
and the results showed that the widespread existence of the strong H$\alpha$ absorption in cool stars, as had been confirmed by observations, indicates the widespread existence of stellar chromospheres.
So the issue of the absence of chromospheres on stars is weakened. 

On the other hand, Vernazza et al. \citep{1973ApJ...184..605V, 1981ApJS...45..635V} modeled the atmosphere of the Sun (from photosphere to chromosphere and transition zone) to understand the structure of the solar chromosphere.
This initial model was later improved by Fontenla et al. and Avrett et al. \citep{1999ApJ...518..480F, 2008ApJS..175..229A}.
The modeling results by the authors showed that the line-core intensity of the H$\alpha$ line is formed in the chromosphere, 
while the central intensities of the H$\beta$ and H$\gamma$ lines have substantial contributions from the photosphere.
The results of the newer model \citep{2008ApJS..175..229A} further show that, 
the photospheric contribution to the central intensity of the H$\gamma$ line is greater than that of the H$\beta$ line for the average quiet-Sun.
Since the physical model and computer program used by the authors can be applied to general stars \citep{2008ApJS..175..229A}, 
the results of these studies are also applicable to solar-like stars in general. 
Then the response of the line-core intensity of the H$\alpha$ line of solar-like stars to stellar magnetic activity would be very different from that of the other Balmer lines (H$\beta$, H$\gamma$, etc.) according to the modeling results.
This issue can be investigated by utilizing a large amount of spectral survey data of stars.

The LAMOST (Large Sky Area Multi-Object Fiber Spectroscopic Telescope) Low-Resolution Spectroscopic Survey (LRS), with the spectral resolution power $\lambda / \Delta\lambda \approx 1800$,
has been collecting the optical spectra of astronomical objects since October 2011 \citep{2012RAA....12.1197C, 2012RAA....12.1243L}.
The wavelength coverage of the LAMOST LRS is about 3700--9000 {\AA} \citep{2012RAA....12.1197C},
so the main part of the hydrogen Balmer series is included in the LRS data, 
which provides an opportunity to study the intensities of the Balmer lines as well as their relations based on a homogeneous dataset \citep{2021RNAAS...5....6H}.
The LAMOST Medium-Resolution Spectroscopic Survey (MRS), starting observations in September 2017, 
has a higher resolution power ($\lambda / \Delta\lambda \approx 7500$) than the LRS.
MRS collects spectral data of astronomical objects in a blue band (about 4950--5350 {\AA}) and a red band (about 6300--6800 {\AA}) \citep{2020arXiv200507210L}. 
The H$\alpha$ line of the Balmer series is included in the red band of MRS,
and other Balmer lines are not covered by the MRS observations.
He et al. \citep{2023Ap&SS.368...63H} investigated the chromospheric activity of F-, G-, and K-type stars using the H$\alpha$ line data obtained by the MRS, 
and the results showed that the H$\alpha$ line can be an effective indicator of stellar chromospheric activity for main-sequence solar-type stars as well as for giant stars. 

\begin{figure*}
  \centering
  \includegraphics[width=1.20\textwidth]{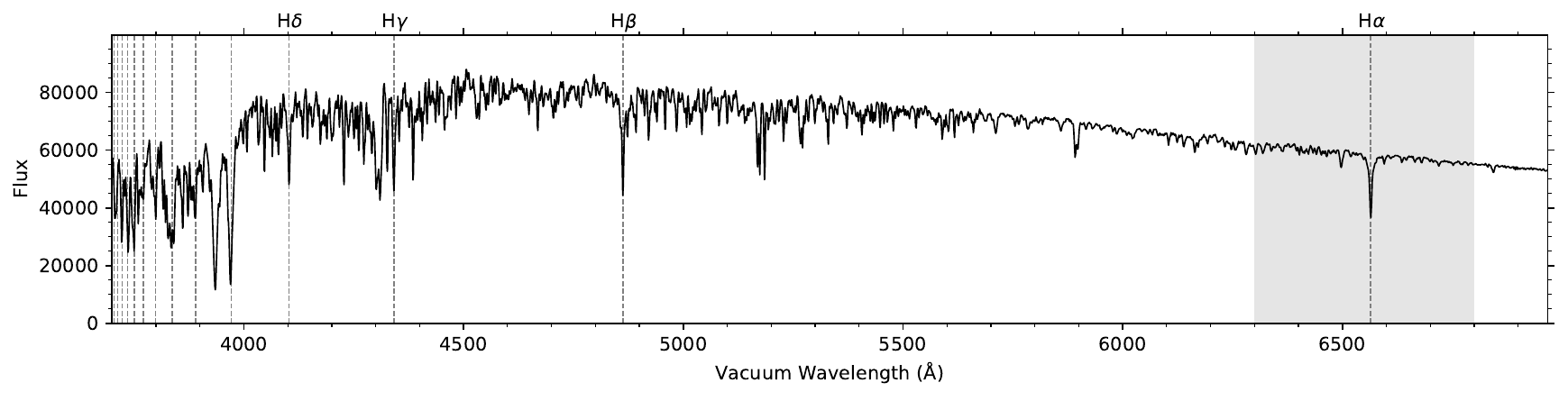}
  \caption{\scriptsize Example of the LAMOST LRS spectra of solar-like stars employed in this work.
    The {\tt\string obsid} (unique LAMOST observation identifier) of the example spectrum is {\tt\string 893207228}, which was observed in February 2021.
    The spectral flux is relatively calibrated (arbitrary units).
    The wavelength values in the plot are in the rest frame.
    The vertical dashed lines indicate the wavelengths of the Balmer lines contained in the LRS spectrum.
    The first four Balmer lines (H$\alpha$, H$\beta$, H$\gamma$, and H$\delta$ lines) analyzed in this work are labeled;
    the higher order Balmer lines are not involved in this work for line blending.
    The shaded area shows the wavelength coverage of the red band of MRS for reference.}
  \label{fig:balmer_lines_in_lrs}
\end{figure*}

In this paper, the intensities of the Balmer lines and their relations are investigated for stellar chromospheric activity of solar-like stars based on the spectral data of the LAMOST spectroscopic surveys.
The spectral data of solar-like stars collected by both the LRS and MRS are used,
so the results of the two surveys can be cross-checked with each other.
The stellar effective temperature of the solar-like spectral sample used in this work is limited to a narrow range around the effective temperature of the Sun. 
Figure~\ref{fig:balmer_lines_in_lrs} shows an example of the LRS spectra of solar-like stars employed in this work.
The wavelengths of the Balmer lines contained in the LRS spectrum are indicated by the vertical dashed lines in Fig.~\ref{fig:balmer_lines_in_lrs}.
The wavelength coverage of the red band of MRS is also displayed in Fig.~\ref{fig:balmer_lines_in_lrs} for reference (see the shaded area). 
In this work, the first to fourth Balmer lines contained in the LRS data are analyzed, 
i.e., the H$\alpha$, H$\beta$, H$\gamma$, and H$\delta$ lines (labeled in Fig.~\ref{fig:balmer_lines_in_lrs}) with the principal quantum number $n=3, 4, 5$, and 6, respectively.
The higher order Balmer lines are not involved in the analysis because they are blended with other spectral lines in the low resolution LRS data as demonstrated in Fig.~\ref{fig:balmer_lines_in_lrs}.
Besides, the higher order Balmer lines are close to the ultraviolet end of the LRS spectrum and generally have a larger uncertainty of spectral fluxes compared with the first four Balmer lines. 
As for the MRS, the H$\alpha$ line data contained in the MRS spectra of solar-like stars co-sourced with the LRS spectra are analyzed. 

The content of the paper is organized as follows. 
Section \ref{sec:data} describes the co-source LRS and MRS spectral samples and the stellar source sample of solar-like stars employed in this work. 
Section \ref{sec:measures} introduces and evaluates the activity measures of the Balmer lines (the activity indexes and the absolute flux indexes) used in this work for indicating the intensity magnitudes of the Balmer lines in response to stellar activity. 
In Section \ref{sec:results}, the distributions of the Balmer line intensities and the relation of the intensities between the Balmer lines are analyzed in detail by using the Balmer line indexes derived in Section \ref{sec:measures}.
Section \ref{sec:discuss} gives further discussions on the results obtained in Section \ref{sec:results}. 
Section \ref{sec:conclusion} is the summary and conclusion.

\section{Data} \label{sec:data}

The spectral and catalog data of the LRS and MRS in the LAMOST Data Release 9 (DR9; http://www.lamost.org/dr9/v2.0/) are used for the analysis, 
which consists of ten years of LAMOST sky survey data (up to June 2021). 
The LRS spectral sample of solar-like stars is selected from the {\tt\string LAMOST LRS Stellar Parameter Catalog of A, F, G and K Stars} of DR9.
This catalog provides the stellar effective temperature ($T_\mathrm{eff}$), surface gravity ($\log g$), metallicity ($\mathrm{[Fe/H]}$), and radial velocity for 6\,921\,466 stellar spectra of LRS,
which are determined by the LAMOST Stellar Parameter Pipeline (LASP) \citep{2015RAA....15.1095L} from the LRS spectral data.
The co-source MRS spectral sample is selected from the {\tt\string LAMOST MRS Parameter Catalog} of DR9, 
which provides the stellar parameters determined by the LASP for 1\,696\,952 MRS spectra.
For a small portion of the spectra in the LRS and MRS catalogs, 
the stellar parameters are not full available (that is, one or more parameters are missing in the catalogs); 
these spectra are not used in the analysis.
After this screening, there are 6\,903\,624 LRS spectra and 1\,512\,625 MRS spectra.

The signal-to-noise ratio (S/N) of the spectral data affects the uncertainty of the spectral fluxes of the Balmer lines as well as the uncertainty of the stellar parameters determined from the spectra \citep{2015RAA....15.1095L},
thus the high-S/N data of the LAMOST are used in the analysis.
For the LRS data, the S/N condition used is $\mathrm{S/N}_r \ge 100$ and $\mathrm{S/N}_g \ge 70$ for the $r$ band and $g$ band of LRS, respectively, 
while for the MRS spectra, a comparable S/N condition of $\mathrm{S/N}_R \ge 100$ and $\mathrm{S/N}_B \ge 70$ is used for the red band and blue band of MRS, respectively, 
considering that the both the ratios of $\mathrm{S/N}_r:\mathrm{S/N}_g$ and $\mathrm{S/N}_R : \mathrm{S/N}_B$ are roughly $10:7$ for the LRS and MRS spectra of solar-like stars (see Supplementary Methods and Supplementary Fig.~S1).
After adding the high-S/N screening condition, there are 1\,470\,079 LRS spectra and 373\,312 MRS spectra left.

When evaluating the activity indexes of the Balmer lines (see Section \ref{sec:measures}), 
the continuum fluxes on the two sides of the lines are employed as a benchmark for estimating the intensity levels at the centers of the lines.
The continuum profile of a stellar spectrum is known to vary with stellar effective temperature,
thus the different definitions for the continuum bands on the two sides of the lines are often adopted for different types of stars in the literature \citep{2010AJ....140.1402H, 2023Ap&SS.368...63H, 2003A&A...403.1077K, 2011A&A...534A..30G}.
The situation is even more complicated if multiple Balmer lines are investigated simultaneously,
since the Balmer lines are distributed in a wide wavelength range from red band to violet band of a stellar spectrum (see Fig.~\ref{fig:balmer_lines_in_lrs}).
For this reason, in this work, the stellar effective temperature of the solar-like spectral sample is limited to a relatively narrow range of $\pm 75$\,K around the effective temperature of the Sun (about 5777\,K \citep{1996A&ARv...7..243C}),
which is $\pm 3$ times the peak position ($\sim 25$\,K) of the effective temperature uncertainty distributions of the selected high-S/N LRS and MRS samples (see Supplementary Methods and Supplementary Fig.~S2);
then the continuum profile of the employed spectra of solar-like stars is similar, 
and a fixed set of continuum band definitions can be used to evaluate the activity indexes of the Balmer lines.
The effect of the stellar surface gravity and metallicity on continuum profile is weaker than the effect of the stellar effective temperature \citep{2000AJ....120.1072S, 2021A&A...649A..97L}; 
so a surface gravity condition of $\log g \ge 3.9$ dex is employed for the solar-like sample,
and no further restriction on [Fe/H] is imposed.
The stellar parameter conditions described above for the solar-like spectral sample are applied to both the LRS and MRS data.
After adding the solar-like constraint, there are 125\,904 LRS spectra and 27\,264 MRS spectra left.

In order to find out the stellar sources observed by both the LRS and MRS, 
the {\tt\string uid} (unique LAMOST source identifier) provided in the LRS and MRS catalogs is utilized.
The LAMOST {\tt\string uid} is generated from the sky coordinates of an observed celestial object (see the data release documentation of LAMOST for the details),
thus the LRS and MRS spectra with the same {\tt\string uid} can be regarded as coming from the same source.
After adding the co-source screening condition, there are 8\,578 LRS spectra and 9\,364 MRS spectra left, which belong to 4\,602 stellar sources.
The numbers of the LRS or MRS co-source spectra are larger than the number of the stellar sources because a stellar object may be observed multiple times at different dates by the LRS or MRS.

In addition to the stellar parameters mentioned above, 
the MRS catalog also provides the values of $v \sin i$ (projected rotational velocity; see the {\tt\string vsini\_lasp} parameter in the catalog) determined by the LASP for the spectra with $v \sin i > 30$ km/s. 
The $v \sin i$ values less than 30 km/s are not provided in the MRS catalog (marked as $-9999.0$).
Because large $v \sin i$ values can lead to an excess increment of the evaluated activity indexes of the Balmer lines \citep{2023Ap&SS.368...63H, 1985ApJ...289..269H, 1991A&A...251..199P}, 
only stellar sources with $v \sin i < 30$ km/s (i.e., stellar sources with {\tt\string vsini\_lasp} = $-9999.0$ in the MRS catalog) are kept, as suggested in the work by He et al. \citep{2023Ap&SS.368...63H}. 

On the other hand, the H$\alpha$ lines of a few LRS and MRS spectra present emission-line profiles (i.e., the central flux of H$\alpha$ line exceeds the continuum flux);
those spectra are removed from the sample,
because for the solar-like stars with the stellar effective temperature within the range adopted in this work,
the central flux of the H$\alpha$ line associated with the stellar magnetic activity is generally below the continuum flux \citep{2023Ap&SS.368...63H}.
Those H$\alpha$ emission lines might come from stellar flares \citep{2022ApJ...928..180W, 2022A&A...663A.140L} or from sources between the observed stars and the telescope, 
such as the galactic nebulae \citep{2021RAA....21...96W, 2022RAA....22g5015W}, which are beyond the scope of this paper.

By applying all the screening conditions for the solar-like sample selection described above,
a total of 3\,677 stellar sources are obtained, which are associated with 6\,562 LRS spectra and 6\,915 MRS spectra.
The dataset of the LRS spectral sample, the MRS spectral sample, and the stellar source sample obtained in this work is available online (web link: https://doi.org/10.5281/zenodo.10848180). 
The descriptions of the dataset can be found in Supplementary Note.

The ranges of the stellar atmospheric parameters of the LRS and MRS samples of solar-like stars obtained in this work are 
$T_\mathrm{eff}$: $5777 \pm 75$\,K (as already shown above), 
$\log g$: $3.9 \sim 4.7$\,dex,
and $\mathrm{[Fe/H]}$: $-1.2 \sim 0.6$\,dex.
A correlation analysis between the stellar atmospheric parameters determined from the LRS and MRS data for the stellar source sample of solar-like stars is given in Supplementary Methods.
If there are multiple measurements of a stellar parameter by the LRS or MRS for a given stellar source, 
the median of the multiple observations is used as the representative value of the parameter of the stellar source in the analysis. 
The results of the correlation analysis show that the stellar parameters determined from the LRS and MRS data are generally consistent with each other (see Supplementary Fig.~S2). 
The uncertainties of the stellar atmospheric parameters ($\delta T_\mathrm{eff}$, $\delta \log g$, and $\delta \mathrm{[Fe/H]}$) determined from the LRS and MRS data are at the similar levels (see also Supplementary Fig.~S2);
the peak positions of the $\delta T_\mathrm{eff}$, $\delta \log g$, and $\delta \mathrm{[Fe/H]}$ distributions of the LRS and MRS data are at about 25\,K (as already shown above), 0.03\,dex, and 0.02\,dex, respectively.

\section{Activity measures of Balmer lines} \label{sec:measures}

A commonly used measure for indicating the intensity level of a stellar chromospheric spectral line in response to stellar magnetic activity is the activity index,
which is defined as the ratio of the flux at line center to the continuum flux around the line;
and the continuum flux, in turn, is estimated through the fluxes in the two predefined continuum bands on the two sides of the line. 
For the different types of stars and the observations with different spectral resolutions, 
there is usually a variation in the shape of the line profile and the shape of the continuum profile,
and hence a variation in the definitions of the central band and the continuum bands of the line. 
Section \ref{sec:index} describes the definitions of the central and continuum bands of the Balmer lines adopted in this work for the LAMOST spectra of solar-like stars,
as well as the formulas used to calculate the activity indexes of the Balmer lines for the LRS and MRS data.

The value of the activity index of a Balmer line depends on the continuum flux around the line; 
and the continuum fluxes of different Balmer lines are generally different.
Thus, the activity index can be utilized to reflect the activity variation of an individual line, 
but may not be appropriate for comparing intensities magnitudes of different lines.
In order to investigate intensities of different Balmer lines, 
on the basis of the derived activity index values, 
the absolute fluxes at the centers of the Balmer lines are further evaluated,
and an absolute flux index is introduced for each of the Balmer lines to indicate the absolute intensity magnitude of the line; 
then the relative intensities between the Balmer lines can be investigated. 
Section \ref{sec:absflux} describes the approaches for deriving the absolute flux indexes of the Balmer lines in the LRS and MRS data.

\subsection{Activity indexes of Balmer lines} \label{sec:index}

\subsubsection{Activity indexes for LRS data} \label{sec:index_LRS}

\begin{figure*}
  \centering
  \includegraphics[width=1.20\textwidth]{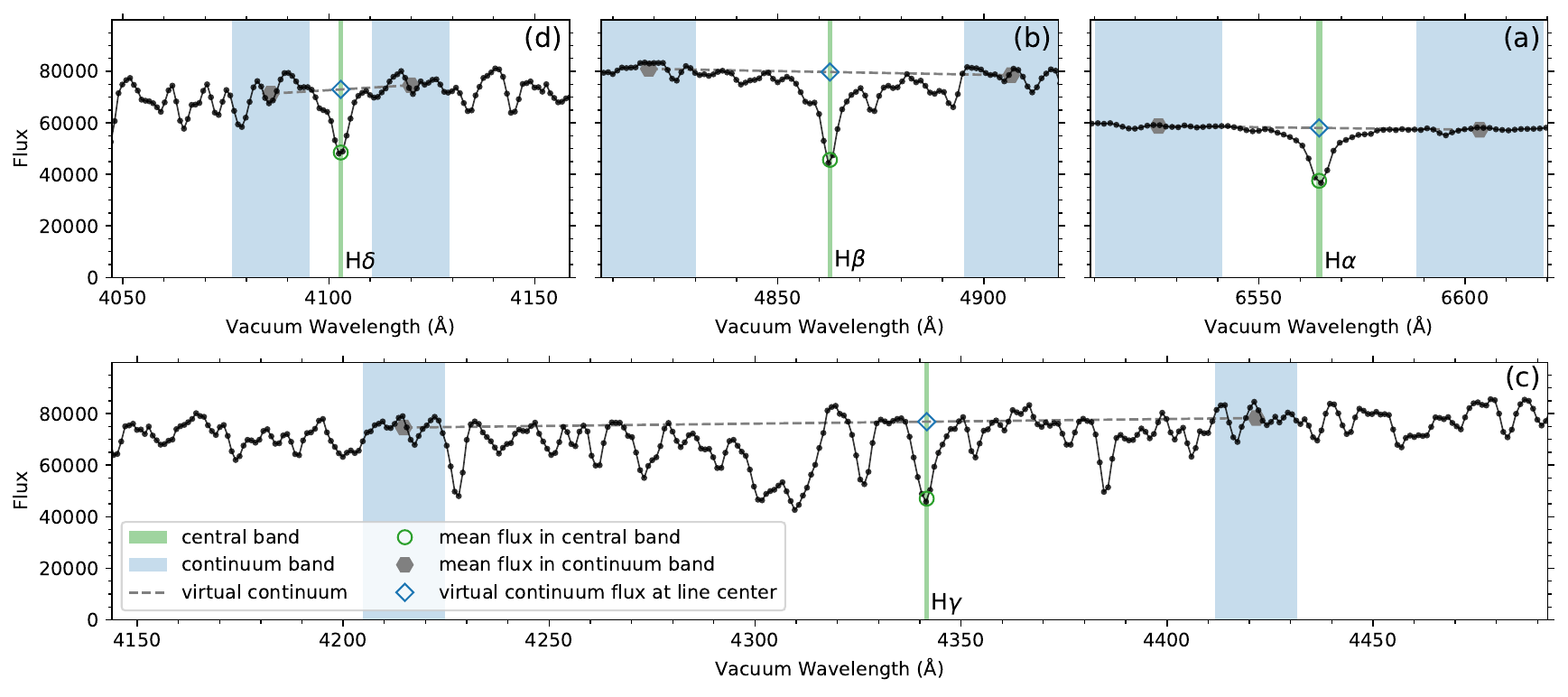}
  \caption{\scriptsize Diagram illustrating the central and continuum bands adopted in this work for deriving the activity indexes of the Balmer (a) H$\alpha$, (b) H$\beta$, (c) H$\gamma$, and (d) H$\delta$ lines, as tabulated in Table~\ref{tab:bands_of_lines}.
  The example LRS spectrum is the same as in Fig.~\ref{fig:balmer_lines_in_lrs}.
  The spectral flux is relatively calibrated (arbitrary units).
  The wavelength values are in the rest frame.
  For each of the Balmer lines, the central band is shown in green and the continuum bands are in blue;
  the filled hexagons indicate the mean fluxes in the two continuum bands, 
  and the dashed line represents the virtual continuum linearly interpolated from the mean fluxes of the two continuum bands;
  the diamond symbol indicates the virtual continuum flux at the center wavelength of the Balmer line,
  and the circle symbol indicates the mean flux in the central band of the Balmer line 
  (see the main text in Section \ref{sec:index} for details).
  }
  \label{fig:lrs_balmer_line_indexes}
\end{figure*}

The activity indexes are evaluated for the first four Balmer lines (H$\alpha$, H$\beta$, H$\gamma$, and H$\delta$ lines) in the LRS data as explained in Section \ref{sec:intro}. 
For each of the lines, a central band at the line center and two continuum bands on the two sides of the line are specified to derive the activity index of the line.
Table~\ref{tab:bands_of_lines} tabulates the center wavelengths of the four Balmer lines (denoted by $\lambda_\mathrm{cen}$),
the center wavelengths of the two continuum bands on the violet side and the red side of the lines (denoted by $\lambda_\mathrm{cont1}$ and $\lambda_\mathrm{cont2}$, respectively), 
and the widths of the central and continuum bands adopted in this work. 
All the wavelength values displayed in Table~\ref{tab:bands_of_lines} are in vacuum as adopted by the LAMOST.
The center wavelengths of the four Balmer lines shown in Table~\ref{tab:bands_of_lines} are from the paper by Stoughton et al. \citep{2002AJ....123..485S}, as used by the Sloan Digital Sky Survey (SDSS).
The locations of the continuum bands are determined by referring to the continuum profile of the LRS spectra of solar-like stars.
The widths of the central and continuum bands are determined by ensuring that each central band contains at lease one data point and each continuum band contains about 20 data points;
this choice of the band widths is necessary to keep the uncertainty of the derived activity indexes at an acceptable level. 
Figures~\ref{fig:lrs_balmer_line_indexes}a--\ref{fig:lrs_balmer_line_indexes}d give a diagram illustration of the central and continuum bands adopted in this work for the H$\alpha$, H$\beta$, H$\gamma$, and H$\delta$ lines, respectively, as listed in Table~\ref{tab:bands_of_lines}.

\begin{table*}
\centering
\caption{\small Central and continuum bands adopted in this work for deriving the activity indexes of the Balmer lines.}
\label{tab:bands_of_lines}
\scriptsize
\begin{tabular}{ccclrlcclrlcc}
\hline\hline
Balmer    & \multicolumn{2}{c}{Central band} && \multicolumn{4}{c}{Continuum band on the violet side}    && \multicolumn{4}{c}{Continuum band on the red side}       \\ 
             \cline{2-3}                         \cline{5-8}                                                   \cline{10-13}  
line      & Center wavelength & Width        && \multicolumn{2}{c}{Center wavelength} & Width   & $a$      && \multicolumn{2}{c}{Center wavelength} & Width   & $b$      \\
          & ({\AA})           & ({\AA})      && \multicolumn{2}{c}{({\AA})}           & ({\AA}) &          && \multicolumn{2}{c}{({\AA})}           & ({\AA}) &          \\
\hline
H$\alpha$ & $6564.61$         & $1.6$        && $6525.61$ & ($-39$)                   & $31$    & $0.5$    && $6603.61$ & ($+39$)                   & $31$    & $0.5$    \\ 
H$\beta$  & $4862.68$         & $1.2$        && $4818.68$ & ($-44$)                   & $23$    & $0.5$    && $4906.68$ & ($+44$)                   & $23$    & $0.5$    \\
H$\gamma$ & $4341.68$         & $1.0$        && $4214.68$ & ($-127$)                  & $20$    & $0.3865$ && $4421.68$ & ($+80$)                   & $20$    & $0.6135$ \\
H$\delta$ & $4102.89$         & $1.0$        && $4085.89$ & ($-17$)                   & $19$    & $0.5$    && $4119.89$ & ($+17$)                   & $19$    & $0.5$    \\
\hline
\multicolumn{12}{p{\textwidth}}{Note:} \\
\multicolumn{12}{p{\textwidth}}{(1) All wavelength values are in vacuum as adopted by the LAMOST.} \\
\multicolumn{12}{p{\textwidth}}{(2) The values in the parentheses for the center wavelengths of the two continuum bands are the wavelength offsets relative to the center wavelengths of the lines.} \\
\multicolumn{12}{p{\textwidth}}{(3) $a$ and $b$ are the weight factors for the fluxes of the two continuum bands used in Equation (\ref{equ:lrs_Lindex}), which are calculated from Equations (\ref{equ:factor_a}) and (\ref{equ:factor_b}), respectively; the sum of $a$ and $b$ equals unity.} \\
\end{tabular}
\end{table*}

For each of the four Balmer lines in the LRS data,
the mean flux in the central band (denoted by $\overline{F}_\mathrm{cen}$; indicated by the circle symbol in Fig.~\ref{fig:lrs_balmer_line_indexes}) and the mean fluxes in the two continuum bands on the violet side and red side of the line (denoted by $\overline{F}_\mathrm{cont1}$ and $\overline{F}_\mathrm{cont2}$, respectively; indicated by the filled hexagons in Fig.~\ref{fig:lrs_balmer_line_indexes}) are measured;
then the activity index of the Balmer line (denoted by $L$, where the letter L indicates that the activity index is derived from LRS data) is calculated by using the following formula:
\begin{equation} \label{equ:lrs_Lindex}
  L = \frac{\overline{F}_\mathrm{cen}}{a \cdot \overline{F}_\mathrm{cont1}+b \cdot \overline{F}_\mathrm{cont2}}.
\end{equation}
The coefficients $a$ and $b$ in Equation (\ref{equ:lrs_Lindex}) are the weight factors for $\overline{F}_\mathrm{cont1}$ and $\overline{F}_\mathrm{cont2}$, respectively, 
whose values are inversely proportional to the distances of the continuum bands from the center of the line, that is,
\begin{equation} \label{equ:factor_a}
  a = \frac{\lambda_\mathrm{cont2}-\lambda_\mathrm{cen}}{\lambda_\mathrm{cont2}-\lambda_\mathrm{cont1}},
\end{equation}
\begin{equation} \label{equ:factor_b}
  b = \frac{\lambda_\mathrm{cen}-\lambda_\mathrm{cont1}}{\lambda_\mathrm{cont2}-\lambda_\mathrm{cont1}},
\end{equation}
and the sum of $a$ and $b$ equals unity.
The specific values of $a$ and $b$ of the four Balmer lines are given in Table~\ref{tab:bands_of_lines}.
If the two continuum bands are symmetrical with respect to the center of the line, both $a$ and $b$ are $0.5$ (i.e., equally weighted).
If the two continuum bands are asymmetrical relative to the line center, such as for the H$\gamma$ line as illustrated in Fig.~\ref{fig:lrs_balmer_line_indexes}c, 
the values of $a$ and $b$ are different (see Table~\ref{tab:bands_of_lines}).
The $a \cdot \overline{F}_\mathrm{cont1}+b \cdot \overline{F}_\mathrm{cont2}$ term in the denominator of Equation (\ref{equ:lrs_Lindex}) can also be understood as the virtual continuum flux at the center wavelength of the Balmer line (indicated by the diamond symbol in Fig.~\ref{fig:lrs_balmer_line_indexes}), 
which is linearly interpolated from the values of $\overline{F}_\mathrm{cont1}$ and $\overline{F}_\mathrm{cont2}$ as implied in Equations (\ref{equ:factor_a}) and (\ref{equ:factor_b}) and illustrated by the dashed line in Fig.~\ref{fig:lrs_balmer_line_indexes}.

The $L$-index values of the four Balmer lines (denoted by $L_\mathrm{H\alpha}$, $L_\mathrm{H\beta}$, $L_\mathrm{H\gamma}$, and $L_\mathrm{H\delta}$, respectively) are derived for the LRS spectral sample of solar-like stars obtained in Section \ref{sec:data}.
The wavelength shift in the original LRS data caused by the stellar radial velocity is corrected by using the {\tt\string rv} parameter in the LRS catalog before the activity index evaluation.
The uncertainties of the activity index values are also estimated by considering the flux uncertainty, the radial velocity uncertainty, and the uncertainty due to the discrete data points \citep{2023Ap&SS.368...63H}.

As explained in Section \ref{sec:data}, a stellar object may be observed multiple times at different dates by the LRS.
Then, for a same stellar source, there might be multiple measurements of the $L$-index for each of the Balmer lines.
These multiple $L$-index measurements of each Balmer line probably correspond to different activity levels of the source and present a certain degree of fluctuations around a mid value.
For this reason, the median of the multiple measurements of each Balmer line is used as the representative $L$-index value of the line of the same source.
This operation is performed for each stellar object in the stellar source sample of solar-like stars obtained in Section \ref{sec:data}.

The data of the $L_\mathrm{H\alpha}$, $L_\mathrm{H\beta}$, $L_\mathrm{H\gamma}$, and $L_\mathrm{H\delta}$ indexes (and their uncertainties) of the four Balmer line of the LRS spectral sample and the stellar source sample are provided in the dataset of this paper (see Supplementary Note and Supplementary Tables~S1 and S3).
Some $L$-index and uncertainty values are unavailable for a small portion of the LRS spectral sample owing to the invalid LRS flux data around the Balmer lines; 
those missing values (as well as the missing values for other activity measures obtained in this work) are marked as $-9999.0$ in the dataset.

\begin{figure*}
  \centering
  \includegraphics[width=0.85\textwidth]{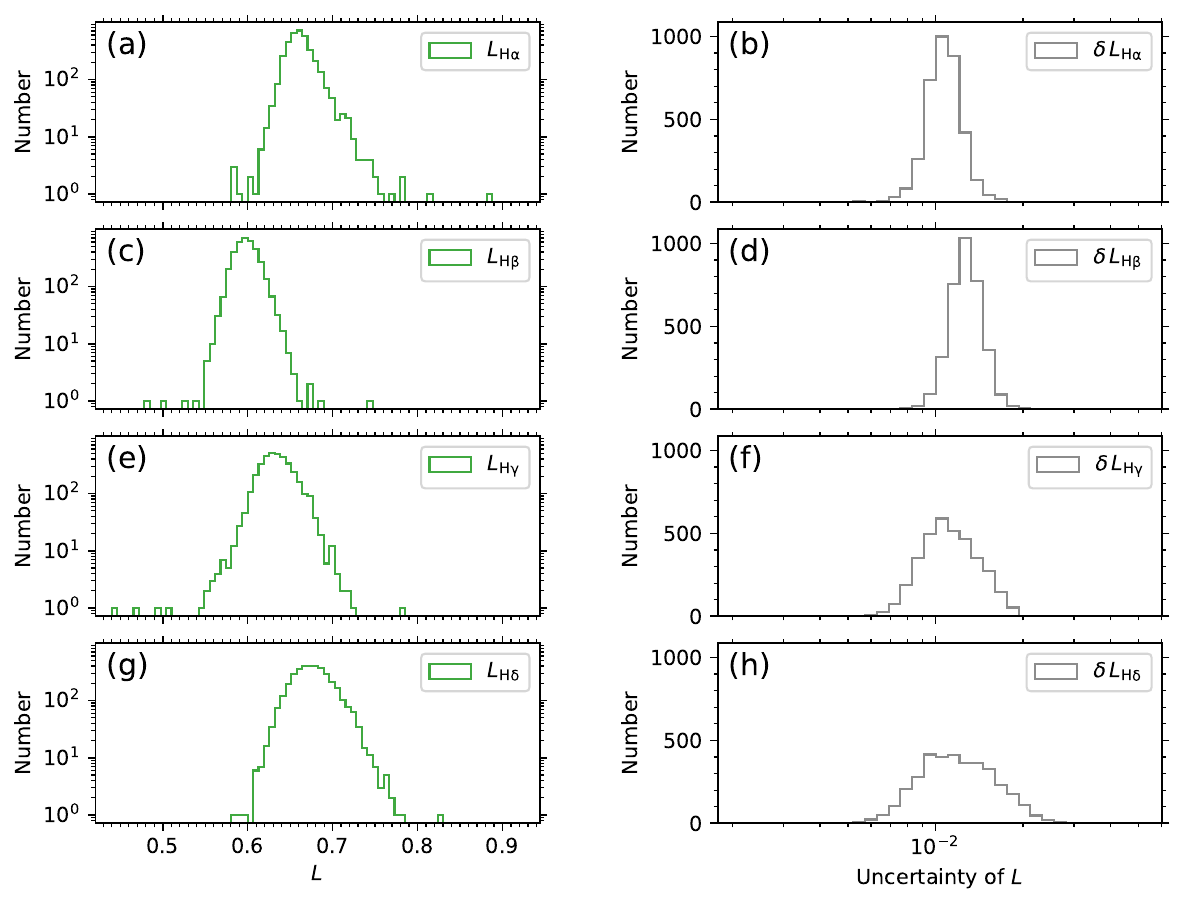}
  \caption{\scriptsize Distribution histograms of the activity indexes $L_\mathrm{H\alpha}$, $L_\mathrm{H\beta}$, $L_\mathrm{H\gamma}$, and $L_\mathrm{H\delta}$ of the four Balmer lines (panels (a), (c), (e), and (g)) and their uncertainties $\delta L_\mathrm{H\alpha}$, $\delta L_\mathrm{H\beta}$, $\delta L_\mathrm{H\gamma}$, and $\delta L_\mathrm{H\delta}$ (panels (b), (d), (f), and (h)) derived from the LRS data for the stellar source sample of solar-like stars employed in this work.
  }
  \label{fig:Lindex_err_histogram}
\end{figure*}

Figure~\ref{fig:Lindex_err_histogram} shows the distribution histograms of the derived $L_\mathrm{H\alpha}$, $L_\mathrm{H\beta}$, $L_\mathrm{H\gamma}$, and $L_\mathrm{H\delta}$ index values of the four Balmer lines (see Figs.~\ref{fig:Lindex_err_histogram}a, \ref{fig:Lindex_err_histogram}c, \ref{fig:Lindex_err_histogram}e, and \ref{fig:Lindex_err_histogram}g) and their uncertainties $\delta L_\mathrm{H\alpha}$, $\delta L_\mathrm{H\beta}$, $\delta L_\mathrm{H\gamma}$, and $\delta L_\mathrm{H\delta}$ (see Figs.~\ref{fig:Lindex_err_histogram}b, \ref{fig:Lindex_err_histogram}d, \ref{fig:Lindex_err_histogram}f, and \ref{fig:Lindex_err_histogram}h) for the stellar source sample of solar-like stars employed in this work.
The medians of the $L$ and $\delta L$ distributions of the four Balmer lines shown in Fig.~\ref{fig:Lindex_err_histogram} are given in Table~\ref{tab:median_of_measures}.
It can be seen from Fig.~\ref{fig:Lindex_err_histogram} and Table~\ref{tab:median_of_measures} that the $L$-index values of the four Balmer lines are in the range of $0.5 \sim 0.9$, 
and the uncertainties of the $L$-index values are on the order of magnitude of $10^{-2}$. 
The relatively wider $\delta L_\mathrm{H\gamma}$ and $\delta L_\mathrm{H\delta}$ distributions (Figs.~\ref{fig:Lindex_err_histogram}f and \ref{fig:Lindex_err_histogram}h) than the $\delta L_\mathrm{H\alpha}$ and $\delta L_\mathrm{H\beta}$ distributions (Figs.~\ref{fig:Lindex_err_histogram}b and \ref{fig:Lindex_err_histogram}d) is owing to the lower S/N values in the blue band than in the red band of LRS. 
The distributions of the $L_\mathrm{H\alpha}$, $L_\mathrm{H\beta}$, $L_\mathrm{H\gamma}$, and $L_\mathrm{H\delta}$ indexes of the four Balmer line shown in Fig.~\ref{fig:Lindex_err_histogram} will be analyzed in detail in Section \ref{sec:results}.

\begin{table*}
\centering
\caption{\small Median values of the activity measures and their uncertainties of the four Balmer lines for the stellar source sample of solar-like stars employed in this work.}
\label{tab:median_of_measures}
\scriptsize
\begin{tabular}{clccccclccccc}
\hline\hline
Balmer && \multicolumn{5}{c}{LRS data} && \multicolumn{5}{c}{MRS data} \\
          \cline{3-7}                     \cline{9-13} 
line   && $L$                  & $\delta L$                  && $\ell$ ({\AA}$^{-1}$)               & $\delta \ell$ ({\AA}$^{-1}$) 
       && $M_\mathrm{H\alpha}$ & $\delta M_\mathrm{H\alpha}$ && $m_\mathrm{H\alpha}$ ({\AA}$^{-1}$) & $\delta m_\mathrm{H\alpha}$ ({\AA}$^{-1}$) \\
\hline
H$\alpha$ && 0.661 & 0.011 && $0.739 \times 10^{-4}$ & $1.2 \times 10^{-6}$ 
          && 0.6594 & 0.0016 && $0.7385 \times 10^{-4}$ & $2.4 \times 10^{-7}$ \\
H$\beta$  && 0.598 & 0.013 && $0.874 \times 10^{-4}$ & $1.9 \times 10^{-6}$  && \ldots & \ldots && \ldots & \ldots \\
H$\gamma$ && 0.633 & 0.011 && $0.860 \times 10^{-4}$ & $1.7 \times 10^{-6}$  && \ldots & \ldots && \ldots & \ldots \\
H$\delta$ && 0.676 & 0.012 && $0.914 \times 10^{-4}$ & $1.8 \times 10^{-6}$  && \ldots & \ldots && \ldots & \ldots \\
\hline
\multicolumn{13}{p{\textwidth}}{Note: Activity measures $L$ (activity index) and $\ell$ (absolute flux index) are for the four Balmer lines in the LRS data;
  activity measures $M_\mathrm{H\alpha}$ (activity index) and $m_\mathrm{H\alpha}$ (absolute flux index) are for the H$\alpha$ line in the MRS data.
  They are defined and evaluated in Section \ref{sec:index} and Section \ref{sec:absflux}.
} \\
\end{tabular}
\end{table*}

\subsubsection{Activity index for MRS data} \label{sec:index_MRS}

The MRS data have a higher spectral resolution than the LRS data, 
but only contain the H$\alpha$ line of the Balmer series (see Fig.~\ref{fig:balmer_lines_in_lrs}).
An activity index, $I_\mathrm{H\alpha}$, was introduced for the H$\alpha$ line in the MRS data in the work by He et al. \citep{2023Ap&SS.368...63H}.
The formula of $I_\mathrm{H\alpha}$ is similar to that of the $L$-index defined for the Balmer lines in the LRS data in Equation (\ref{equ:lrs_Lindex}),
but the definition of the central and continuum bands of the $I_\mathrm{H\alpha}$ index is different from that of the $L_\mathrm{H\alpha}$ index because of the higher spectral resolution of the MRS data.
In the definition of the $I_\mathrm{H\alpha}$ index, the width of the central band of H$\alpha$ line is 0.25\,{\AA};
the wavelength offset of the two continuum bands is $\pm 25$\,{\AA} relative to the H$\alpha$ line center;
and the width of each continuum band is 5\,{\AA} \citep{2023Ap&SS.368...63H}.

Because the MRS and LRS data are collected in different spectral resolutions, 
for a same stellar source, the $I_\mathrm{H\alpha}$ value derived from the MRS data and the $L_\mathrm{H\alpha}$ value from the LRS data are generally different.
In order to obtain a MRS H$\alpha$ activity index that is compatible with the LRS $L_\mathrm{H\alpha}$ index,
the spectral resolution of the original MRS data is first reduced to the same resolution as the LRS data by Gauss smoothing,
then the H$\alpha$ activity index is calculated from the smoothed MRS data using the same approach as for the $L_\mathrm{H\alpha}$ index of the LRS data.
The obtained H$\alpha$ activity index based on the smoothed MRS data is denoted by $\widetilde{I}_\mathrm{H\alpha}$,
and the formula for calculating $\widetilde{I}_\mathrm{H\alpha}$ is 
\begin{equation} \label{equ:I_tilde_index}
  \widetilde{I}_\mathrm{H\alpha} = \frac{\overline{\mathcal{F}}_\mathrm{cen,H\alpha}}
  {a_\mathrm{H\alpha} \cdot \overline{\mathcal{F}}_\mathrm{cont1,H\alpha} + b_\mathrm{H\alpha} \cdot \overline{\mathcal{F}}_\mathrm{cont2,H\alpha}},
\end{equation}
where $\mathcal{F}$ represents the smoothed MRS spectral flux.
The definitions for the central and continuum bands in Equation (\ref{equ:I_tilde_index}) and the weight factors $a_\mathrm{H\alpha}$ and $b_\mathrm{H\alpha}$ for the mean fluxes of the two continuum bands are the same as listed in Table~\ref{tab:bands_of_lines} (note that both $a_\mathrm{H\alpha}$ and $b_\mathrm{H\alpha}$ equal to $0.5$). 

The values of the $\widetilde{I}_\mathrm{H\alpha}$ index are derived for the MRS spectral sample of solar-like stars obtained in Section \ref{sec:data}.
The wavelength shift in the original MRS data caused by the stellar radial velocity is corrected by using the {\tt\string rv\_r0} parameter in the MRS catalog before the $\widetilde{I}_\mathrm{H\alpha}$ index evaluation.
The uncertainties of the $\widetilde{I}_\mathrm{H\alpha}$ values are also estimated using the similar approach as for the $L$-index of the LRS spectra.

A stellar object may be observed multiple times at different dates by the MRS (see Section \ref{sec:data}),
and hence there might be multiple measurements of the $\widetilde{I}_\mathrm{H\alpha}$ index for a same stellar source.
As the operation for the multiple measurements of the $L$-index of the LRS data, 
for each stellar object in the stellar source sample of solar-like stars obtained in Section \ref{sec:data},
the multiple measurements of the $\widetilde{I}_\mathrm{H\alpha}$ index of the object are aggregated by the median of the multiple $\widetilde{I}_\mathrm{H\alpha}$ values.

\begin{figure*}
  \centering
  \includegraphics[width=1.0\textwidth]{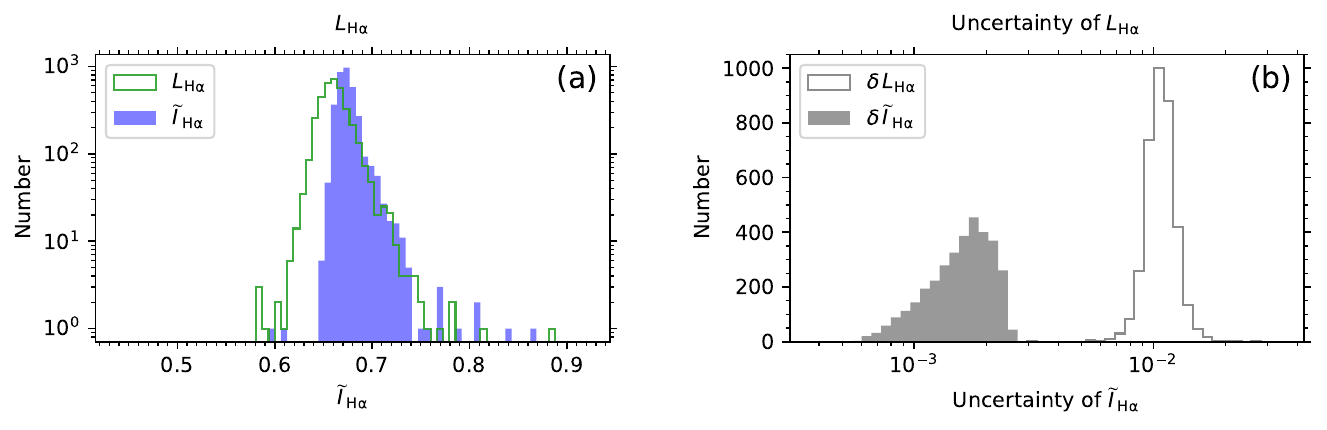}
  \caption{\scriptsize Distribution histograms of the $\widetilde{I}_\mathrm{H\alpha}$ index values (filled histogram in panel (a)) and their uncertainties ($\delta \widetilde{I}_\mathrm{H\alpha}$; filled histogram in panel (b)) derived from the MRS data for the stellar source sample of solar-like stars employed in this work.
    The histograms of the $L_\mathrm{H\alpha}$ index values and their uncertainties ($\delta L_\mathrm{H\alpha}$) of the LRS data for the stellar source sample (line histograms) are also displayed for reference.}
  \label{fig:Itindex_err_histogram}
\end{figure*}

Figure~\ref{fig:Itindex_err_histogram} shows the distribution histograms of the $\widetilde{I}_\mathrm{H\alpha}$ index values (filled histogram in Fig.~\ref{fig:Itindex_err_histogram}a) and their uncertainties $\delta \widetilde{I}_\mathrm{H\alpha}$ (filled histogram in Fig.~\ref{fig:Itindex_err_histogram}b) derived from the MRS data for the stellar source sample of solar-like stars employed in this work.
The histograms of the $L_\mathrm{H\alpha}$ index and its uncertainty $\delta L_\mathrm{H\alpha}$ of the LRS data for the stellar source sample (as having been shown in Figs.~\ref{fig:Lindex_err_histogram}a and \ref{fig:Lindex_err_histogram}b) are also plotted in Fig.~\ref{fig:Itindex_err_histogram} (line histograms) for reference.
It can be seen from Fig.~\ref{fig:Itindex_err_histogram}a that the magnitudes of the $\widetilde{I}_\mathrm{H\alpha}$ index from the MRS data and the $L_\mathrm{H\alpha}$ index from the LRS data are similar.
Figure~\ref{fig:Itindex_err_histogram}b shows that the uncertainty of $\widetilde{I}_\mathrm{H\alpha}$ is about one order of magnitude smaller than the uncertainty of $L_\mathrm{H\alpha}$,
which can be interpreted as the effect of smoothing on the MRS data,
that is, both the spectral flux uncertainty and the uncertainty owing to discrete data points are dramatically reduced by the smoothing procedure.

\begin{figure*}
  \centering
  \includegraphics[width=1.0\textwidth]{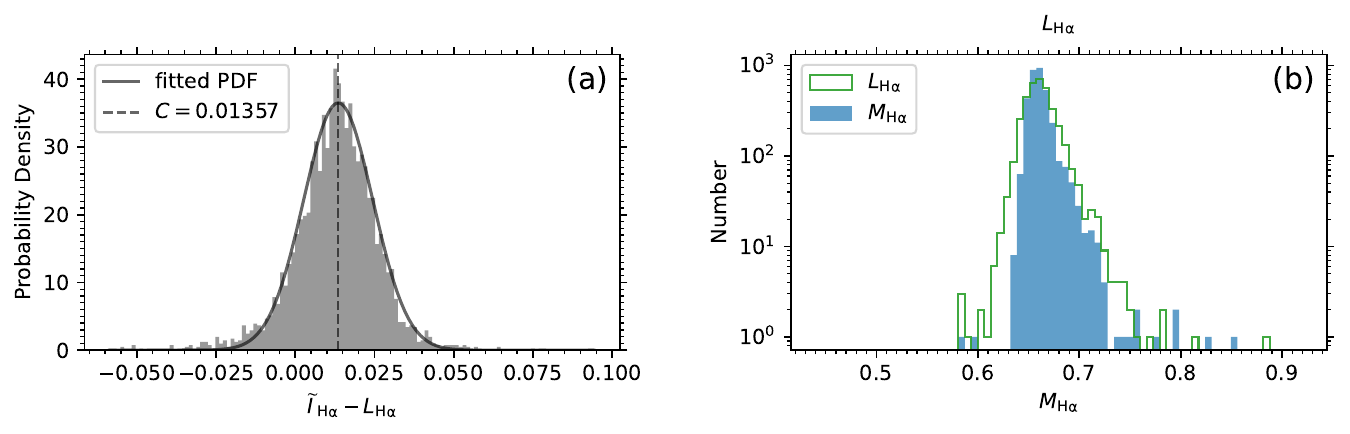}
  \caption{\scriptsize (a) Distribution histogram of the $\widetilde{I}_\mathrm{H\alpha} - L_\mathrm{H\alpha}$ values (difference between $\widetilde{I}_\mathrm{H\alpha}$ and $L_\mathrm{H\alpha}$) for the stellar source sample of solar-like stars employed in this work.
  The solid black curve is the fitted probability density function (PDF) to the envelope of the histogram by a normal distribution model.
  The vertical dashed line indicates the center location (denoted by $C=0.01357$) of the fitted normal distribution. 
  (b) Distribution histogram of the $M_\mathrm{H\alpha}$ index values derived from the MRS data for the stellar source sample of solar-like stars employed in this work (filled histogram).
    The histogram of the $L_\mathrm{H\alpha}$ index values of the LRS data for the stellar source sample (line histogram) is also displayed for reference.} 
  \label{fig:Mindex_offset_histogram}
\end{figure*}

In order to determine the quantitative relationship between the $\widetilde{I}_\mathrm{H\alpha}$ index and the $L_\mathrm{H\alpha}$ index,
the distribution of the difference between the $\widetilde{I}_\mathrm{H\alpha}$ and $L_\mathrm{H\alpha}$ values (i.e., $\widetilde{I}_\mathrm{H\alpha} - L_\mathrm{H\alpha}$) is investigated for the stellar source sample of solar-like stars employed in this work,
which is shown in Fig.~\ref{fig:Mindex_offset_histogram}a.
It can be seen from Fig.~\ref{fig:Mindex_offset_histogram}a that there is a small offset between the values of $\widetilde{I}_\mathrm{H\alpha}$ and $L_\mathrm{H\alpha}$.
To find out the exact value of the offset, the envelope of the $\widetilde{I}_\mathrm{H\alpha} - L_\mathrm{H\alpha}$ distribution histogram shown in Fig.~\ref{fig:Mindex_offset_histogram}a is fitted by a normal distribution model. 
The fitted probability density function (PDF) of the normal distribution is also plotted in Fig.~\ref{fig:Mindex_offset_histogram}a as a solid black curve.  
The center location of the fitted normal distribution (denoted by $C=0.01357$ and indicated by a vertical dashed line in Fig.~\ref{fig:Mindex_offset_histogram}a) reflects the offset between $\widetilde{I}_\mathrm{H\alpha}$ and $L_\mathrm{H\alpha}$,
and the standard deviation of the fitted normal distribution (equaling to $0.01093$) has the similar scale as the $L_\mathrm{H\alpha}$ uncertainty (about $10^{-2}$ with a median of 0.011; see Fig.~\ref{fig:Lindex_err_histogram}b and Table~\ref{tab:median_of_measures}), 
which implies that the scatter of the $\widetilde{I}_\mathrm{H\alpha} - L_\mathrm{H\alpha}$ distribution is mainly caused by the uncertainty of $L_\mathrm{H\alpha}$,
since the uncertainty of $\widetilde{I}_\mathrm{H\alpha}$ is about one order of magnitude smaller than that of $L_\mathrm{H\alpha}$ (see Fig.~\ref{fig:Itindex_err_histogram}b).

To compensate for the offset between $\widetilde{I}_\mathrm{H\alpha}$ and $L_\mathrm{H\alpha}$,
a new activity index (denoted by $M_\mathrm{H\alpha}$, where the letter M indicates that the index is derived from the MRS data) is introduced based on the $\widetilde{I}_\mathrm{H\alpha}$ index, 
which is calculated with the formula
\begin{equation} \label{equ:M_index}
  M_\mathrm{H\alpha} = \widetilde{I}_\mathrm{H\alpha} - 0.01357,
\end{equation}
where the constant number $0.01357$ reflects the offset between $\widetilde{I}_\mathrm{H\alpha}$ and $L_\mathrm{H\alpha}$ (see Fig.~\ref{fig:Mindex_offset_histogram}a).
The uncertainty of the $M_\mathrm{H\alpha}$ index ($\delta M_\mathrm{H\alpha}$) is the same as that of the $\widetilde{I}_\mathrm{H\alpha}$ index (i.e., $\delta M_\mathrm{H\alpha} = \delta \widetilde{I}_\mathrm{H\alpha}$) according to the error propagation rules.

The values of the $M_\mathrm{H\alpha}$ index and their uncertainties are derived from the values of the $\widetilde{I}_\mathrm{H\alpha}$ index for the MRS spectral sample and the stellar source sample of solar-like stars obtained in Section \ref{sec:data}.
The data of the $M_\mathrm{H\alpha}$ index (and its uncertainty) are provided in the dataset of this paper (see Supplementary Note and Supplementary Tables~S2 and S3).
Considering that the $\widetilde{I}_\mathrm{H\alpha}$ index is an intermediate parameter,
it is not included in the dataset;
if desired, it can be derived from the data of $M_\mathrm{H\alpha}$ by using Equation (\ref{equ:M_index}).

Figure~\ref{fig:Mindex_offset_histogram}b shows the distribution histogram of the derived $M_\mathrm{H\alpha}$ index values (filled histogram) for the stellar source sample of solar-like stars employed in this work.
The histogram of the $L_\mathrm{H\alpha}$ index for the stellar source sample (as having been shown in Fig.~\ref{fig:Lindex_err_histogram}a) is also plotted in Fig.~\ref{fig:Mindex_offset_histogram}b (line histogram) for reference.
The median of the $M_\mathrm{H\alpha}$ distribution shown in Fig.~\ref{fig:Mindex_offset_histogram}b,
as well as the median of the $\delta M_\mathrm{H\alpha}$ distribution (same as the $\delta \widetilde{I}_\mathrm{H\alpha}$ distribution shown in Fig.~\ref{fig:Itindex_err_histogram}b),
is given in Table~\ref{tab:median_of_measures}.
It can be seen from Fig.~\ref{fig:Mindex_offset_histogram}b that the distribution of the $M_\mathrm{H\alpha}$ index derived from the MRS data is consistent with the distribution of the $L_\mathrm{H\alpha}$ index from the co-source LRS data.
The wider span of the $L_\mathrm{H\alpha}$ index distribution is due to the larger uncertainty of the $L_\mathrm{H\alpha}$ index values.
Detailed analysis on the distributions of the activity indexes will be performed in Section \ref{sec:results}.

\subsection{Absolute flux indexes of Balmer lines} \label{sec:absflux}

\subsubsection{Absolute flux indexes for LRS data} \label{sec:absflux_LRS}

The $L$-index defined for the Balmer lines in the LRS data (see Equation (\ref{equ:lrs_Lindex})) can also be expressed with the stellar absolute flux:
\begin{equation} \label{equ:lrs_index_absflux}
  L = \frac{\overline{f}_\mathrm{cen}}{a \cdot \overline{f}_\mathrm{cont1}+b \cdot \overline{f}_\mathrm{cont2}},
\end{equation}
where $f$ represents the absolute spectral flux on the stellar surface (in unit of erg\ s$^{-1}$\,cm$^{-2}$\,{\AA}$^{-1}$) and $\overline{f}$ denotes the mean flux of $f$ in a given band.
In Section \ref{sec:index}, the $L$-index values have been obtained from the observed LRS spectral data,
then the absolute flux in the central band of a Balmer line, $\overline{f}_\mathrm{cen}$, can be calculated from the $L$-index by using the following equation: 
\begin{equation} \label{equ:lrs_absflux}
  \overline{f}_\mathrm{cen} = (a \cdot \overline{f}_\mathrm{cont1}+b \cdot \overline{f}_\mathrm{cont2}) \cdot L.
\end{equation}

By expressing the absolute flux $\overline{f}_\mathrm{cen}$ as the ratio to the stellar bolometric flux $\sigma T_\mathrm{eff}^4$ (where $\sigma$ is the Stefan-Boltzmann constant), 
the absolute flux index (denoted by $\ell$) is defined for the Balmer lines in the LRS data as
\begin{equation} \label{equ:l_index}
  \ell = \frac{\overline{f}_\mathrm{cen}}{\sigma T_\mathrm{eff}^4} 
  = \frac{a \cdot \overline{f}_\mathrm{cont1} + b \cdot \overline{f}_\mathrm{cont2}}{\sigma T_\mathrm{eff}^4} \cdot L.
\end{equation}
Because $f$ in the numerator of Equation (\ref{equ:l_index}) is the spectral flux (flux per unit wavelength) and the $\sigma T_\mathrm{eff}^4$ term (stellar bolometric flux) in the denominator of Equation (\ref{equ:l_index}) is the integral flux of all wavelengths, the unit of the absolute flux index $\ell$ is {\AA}$^{-1}$.

Equation (\ref{equ:l_index}) is used to transform the $L$-index of the Balmer lines in the LRS data to the absolute flux index $\ell$.
The $\overline{f}_\mathrm{cont1}$ and $\overline{f}_\mathrm{cont2}$ (absolute fluxes of the two continuum bands) and then the 
$(a \cdot \overline{f}_\mathrm{cont1} + b \cdot \overline{f}_\mathrm{cont2})/(\sigma T_\mathrm{eff}^4)$
term in Equation (\ref{equ:l_index}) are calculated by using the synthetic spectra of the PHOENIX stellar atmosphere model \citep{2016sf2a.conf..223A} published in the work by Husser et al. \citep{2013A&A...553A...6H}. 
A comparison between the PHOENIX synthetic spectra and the observed spectra in the literature \citep{2021A&A...649A..97L} showed that the PHOENIX model can provide a satisfactory representation of the continuum flux of solar-like stars.
The stellar atmospheric parameters of the PHOENIX grid of synthetic spectra \citep{2013A&A...553A...6H} utilized in this work are $T_\mathrm{eff}= 5700, 5800, 5900$ K, 
$\log g= 3.5, 4.0, 4.5, 5.0$ dex, and
$\mathrm{[Fe/H]}= -1.5, -1.0, -0.5, 0.0, 0.5, 1.0$ dex,
which cover the stellar atmospheric parameter ranges of the LAMOST spectral samples of solar-like stars employed in this work (see Section \ref{sec:data}).
The spectral resolution of the original PHOENIX synthetic spectra is reduced to the same resolution as the LRS data by Gauss smoothing.
The values of the 
$(a \cdot \overline{f}_\mathrm{cont1} + b \cdot \overline{f}_\mathrm{cont2}) / (\sigma T_\mathrm{eff}^4)$
term in Equation (\ref{equ:l_index}) on the PHOENIX grid are first calculated based on the smoothed PHOENIX synthetic spectra;
the values of 
$(a \cdot \overline{f}_\mathrm{cont1} + b \cdot \overline{f}_\mathrm{cont2}) / (\sigma T_\mathrm{eff}^4)$
between the PHOENIX grid points are then calculated using linear interpolation from the values on the grid.

The $\ell$-index values of the four Balmer lines (denoted by $\ell_\mathrm{H\alpha}$, $\ell_\mathrm{H\beta}$, $\ell_\mathrm{H\gamma}$, and $\ell_\mathrm{H\delta}$, respectively) are calculated from the values of the $L_\mathrm{H\alpha}$, $L_\mathrm{H\beta}$, $L_\mathrm{H\gamma}$, and $L_\mathrm{H\delta}$ indexes obtained in Section \ref{sec:index}.
The uncertainties of the $\ell$-index values are also estimated, 
which originate from the uncertainties of the $L$-index values and the uncertainties of the stellar atmospheric parameters determined from the LRS data.
In order to be consistent with the absolute flux index defined for the MRS data (see the following subsection),
the $\ell$-index values of the four Balmer lines and their uncertainties are only evaluated for the stellar source sample of solar-like stars obtained in Section \ref{sec:data}.
The data of the $\ell_\mathrm{H\alpha}$, $\ell_\mathrm{H\beta}$, $\ell_\mathrm{H\gamma}$, and $\ell_\mathrm{H\delta}$ indexes (and their uncertainties) of the stellar source sample are provided in the dataset of this paper (see Supplementary Note and Supplementary Table~S3).

Figure~\ref{fig:lrs_lindex_err_histogram} shows the distribution histograms of the derived $\ell_\mathrm{H\alpha}$, $\ell_\mathrm{H\beta}$, $\ell_\mathrm{H\gamma}$, and $\ell_\mathrm{H\delta}$ index values of the four Balmer lines (see Figs.~\ref{fig:lrs_lindex_err_histogram}a, \ref{fig:lrs_lindex_err_histogram}c, \ref{fig:lrs_lindex_err_histogram}e, and \ref{fig:lrs_lindex_err_histogram}g) and their uncertainties $\delta \ell_\mathrm{H\alpha}$, $\delta \ell_\mathrm{H\beta}$, $\delta \ell_\mathrm{H\gamma}$, and $\delta \ell_\mathrm{H\delta}$ (see Figs.~\ref{fig:lrs_lindex_err_histogram}b, \ref{fig:lrs_lindex_err_histogram}d, \ref{fig:lrs_lindex_err_histogram}f, and \ref{fig:lrs_lindex_err_histogram}h) for the stellar source sample of solar-like stars employed in this work.
The medians of the $\ell$ and $\delta \ell$ distributions of the four Balmer lines shown in Fig.~\ref{fig:lrs_lindex_err_histogram} are given in Table~\ref{tab:median_of_measures}.
It can be seen from Fig.~\ref{fig:lrs_lindex_err_histogram} and Table~\ref{tab:median_of_measures} that the $\ell$-index values of the four Balmer lines are in the range of $0.6 \times 10^{-4} \sim 1.1 \times 10^{-4}$~{\AA}$^{-1}$, 
and the uncertainties of the $\ell$-index values are on the order of magnitude of $10^{-6}$~{\AA}$^{-1}$. 
The distributions of the $\ell_\mathrm{H\alpha}$, $\ell_\mathrm{H\beta}$, $\ell_\mathrm{H\gamma}$, and $\ell_\mathrm{H\delta}$ indexes of the four Balmer line shown in Fig.~\ref{fig:lrs_lindex_err_histogram} will be analyzed in detail in Section \ref{sec:results}.

\begin{figure*}
  \centering
  \includegraphics[width=0.85\textwidth]{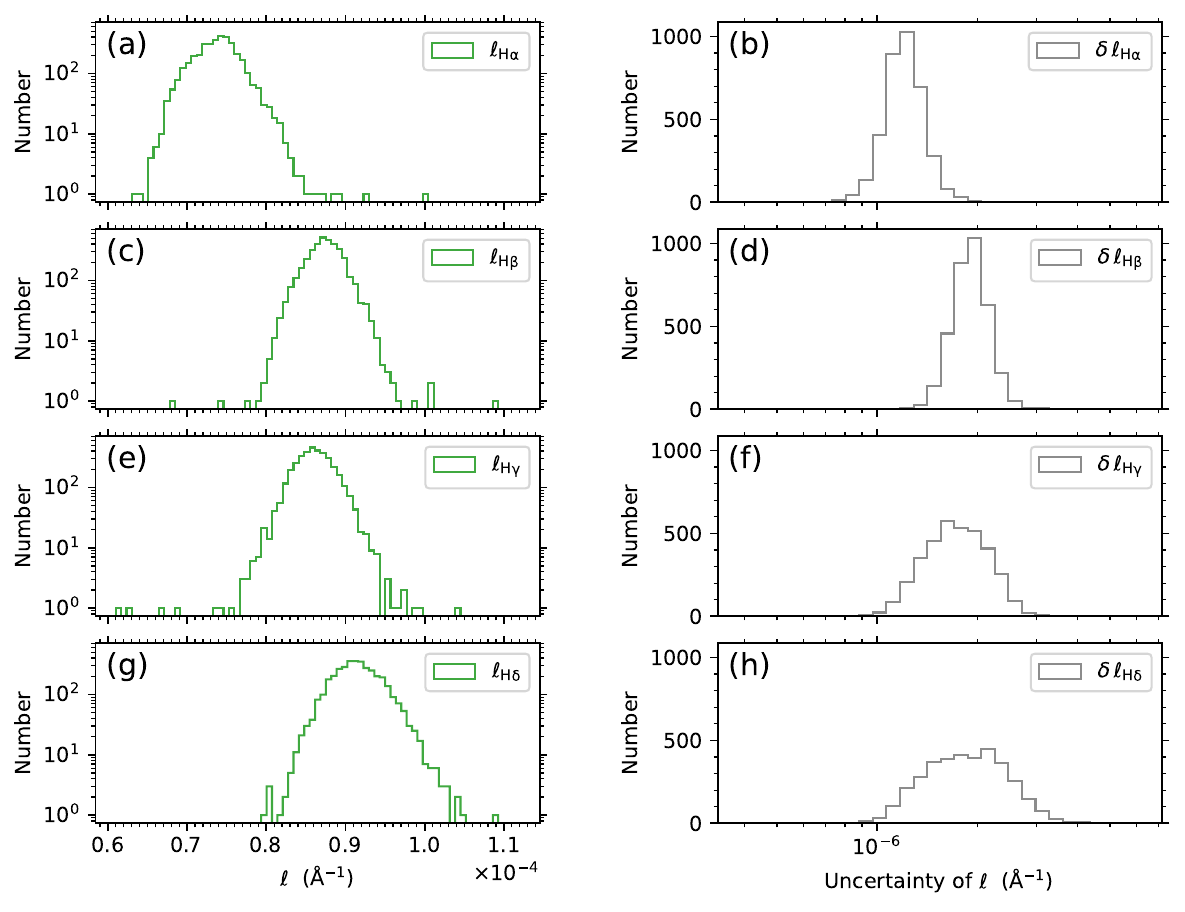}
  \caption{\scriptsize Distribution histograms of the absolute flux indexes $\ell_\mathrm{H\alpha}$, $\ell_\mathrm{H\beta}$, $\ell_\mathrm{H\gamma}$, and $\ell_\mathrm{H\delta}$ of the four Balmer lines (panels (a), (c), (e), and (g)) and their uncertainties $\delta \ell_\mathrm{H\alpha}$, $\delta \ell_\mathrm{H\beta}$, $\delta \ell_\mathrm{H\gamma}$, and $\delta \ell_\mathrm{H\delta}$ (panels (b), (d), (f), and (h)) derived from the LRS data for the stellar source sample of solar-like stars employed in this work.}
  \label{fig:lrs_lindex_err_histogram}
\end{figure*}

\subsubsection{Absolute flux index for MRS data} \label{sec:absflux_MRS}

In Section \ref{sec:index}, the activity index $M_\mathrm{H\alpha}$ have been obtained from the smoothed MRS data,
which is introduced to be quantitatively consistent with the $L_\mathrm{H\alpha}$ index of the LRS data.
Similar to the absolute flux index $\ell$ defined for the Balmer lines in the LRS data, 
an absolute flux index, $m_\mathrm{H\alpha}$, can be defined based on the $M_\mathrm{H\alpha}$ index for the H$\alpha$ line in the MRS data as
\begin{equation} \label{equ:m_index}
  m_\mathrm{H\alpha} = 
  \frac{\overline{f}_\mathrm{cen,H\alpha}}{\sigma T_\mathrm{eff}^4} = 
  \frac{a_\mathrm{H\alpha} \cdot \overline{f}_\mathrm{cont1,H\alpha} + b_\mathrm{H\alpha} \cdot \overline{f}_\mathrm{cont2,H\alpha}}{\sigma T_\mathrm{eff}^4} \cdot M_\mathrm{H\alpha},
\end{equation}
where $a_\mathrm{H\alpha}=b_\mathrm{H\alpha}=0.5$ as usual (see Table~\ref{tab:bands_of_lines}).

In order to be quantitatively consistent with the $\ell_\mathrm{H\alpha}$ index of the LRS data,
the stellar atmospheric parameters determined from the LRS spectra are used to evaluate the values of the 
$(a_\mathrm{H\alpha} \cdot \overline{f}_\mathrm{cont1,H\alpha} + b_\mathrm{H\alpha} \cdot \overline{f}_\mathrm{cont2,H\alpha})/(\sigma T_\mathrm{eff}^4)$
term in Equation (\ref{equ:m_index});
another consideration is that, 
although the uncertainties of the stellar atmospheric parameters of the LRS and MRS data are at the similar level, 
the trailing of the stellar parameter uncertainty distributions of the MRS data is longer than that of the LRS data (see Supplementary Methods and Supplementary Fig.~S2).
For this reason, the $m_\mathrm{H\alpha}$ index and its uncertainty are only evaluated for the stellar source sample of solar-like stars obtained in Section \ref{sec:data}.
The data of the $m_\mathrm{H\alpha}$ index (and its uncertainty) of the stellar source sample are provided in the dataset of this paper (see Supplementary Note and Supplementary Table~S3).

Figure~\ref{fig:mrs_mindex_err_histogram} shows the distribution histograms of the derived $m_\mathrm{H\alpha}$ index values (filled histogram in Fig.~\ref{fig:mrs_mindex_err_histogram}a) and their uncertainties $\delta m_\mathrm{H\alpha}$ (filled histogram in Fig.~\ref{fig:mrs_mindex_err_histogram}b) for the stellar source sample of solar-like stars employed in this work.
The histograms of the $\ell_\mathrm{H\alpha}$ index values and their uncertainty for the stellar source sample (as having been shown in Figs.~\ref{fig:lrs_lindex_err_histogram}a and \ref{fig:lrs_lindex_err_histogram}b) are also plotted in Fig.~\ref{fig:mrs_mindex_err_histogram} (line histograms) for reference. 
The medians of the $m_\mathrm{H\alpha}$ and $\delta m_\mathrm{H\alpha}$ distributions shown in Fig.~\ref{fig:mrs_mindex_err_histogram} are given in Table~\ref{tab:median_of_measures}.
It can be seen from Fig.~\ref{fig:mrs_mindex_err_histogram}a that the distribution of the $m_\mathrm{H\alpha}$ index derived from the MRS data is consistent with the distribution of the $\ell_\mathrm{H\alpha}$ index from the co-source LRS data.
Figure~\ref{fig:mrs_mindex_err_histogram}b shows that the uncertainty of $m_\mathrm{H\alpha}$ is about one order of magnitude smaller than that of $\ell_\mathrm{H\alpha}$.
The wider span of the $\ell_\mathrm{H\alpha}$ index distribution than the $m_\mathrm{H\alpha}$ index distribution (see Fig.~\ref{fig:mrs_mindex_err_histogram}a) is owing to the larger uncertainty of the $\ell_\mathrm{H\alpha}$ index values.
Detailed analysis on the distribution of the absolute flux indexes will be performed in Section~\ref{sec:results}.

\begin{figure*}
  \centering
  \includegraphics[width=1.0\textwidth]{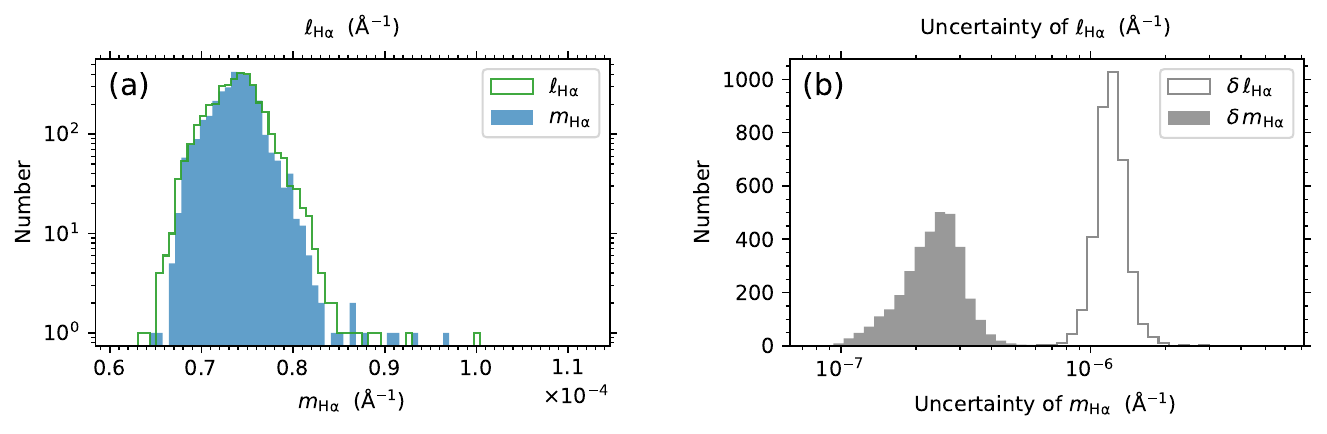}
  \caption{\scriptsize Distribution histograms of the $m_\mathrm{H\alpha}$ index values (filled histogram in panel (a)) and their uncertainties (filled histogram in panel (b)) derived from the MRS data for the stellar source sample of solar-like stars employed in this work.
    The histograms of the $\ell_\mathrm{H\alpha}$ index values and their uncertainties of the LRS data for the stellar source sample (line histograms) are also displayed for reference.}
  \label{fig:mrs_mindex_err_histogram}
\end{figure*}

\section{Results} \label{sec:results}

\subsection{Distribution of the activity indexes of Balmer lines} \label{sec:result_index}

\begin{figure*}
  \centering
  \includegraphics[width=1.01\textwidth]{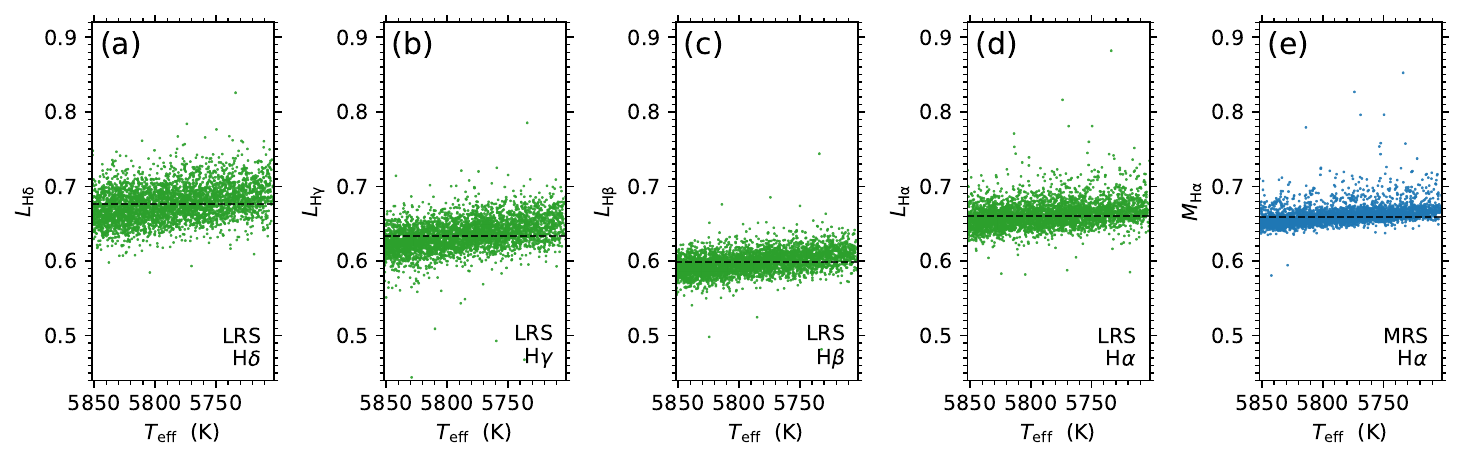}
  \caption{\scriptsize
  Distributions of the activity indexes $L_\mathrm{H\alpha}$, $L_\mathrm{H\beta}$, $L_\mathrm{H\gamma}$, and $L_\mathrm{H\delta}$ (from LRS data) and $M_\mathrm{H\alpha}$ (from MRS data) with stellar effective temperature ($T_\mathrm{eff}$) for the stellar source sample of solar-like stars employed in this work.  
  The horizontal dashed line in each panel indicates the median of the activity index distribution.
  The $T_\mathrm{eff}$ parameter used in the plots is determined from the LRS data.   
  }
  \label{fig:LMindex_with_teff}
\end{figure*}

In Section \ref{sec:index}, the activity indexes $L_\mathrm{H\alpha}$, $L_\mathrm{H\beta}$, $L_\mathrm{H\gamma}$, and $L_\mathrm{H\delta}$ from the LRS data and $M_\mathrm{H\alpha}$ from the MRS data have been obtained for the co-source sample of solar-like stars.
The distribution histograms of the activity indexes (see Figs.~\ref{fig:Lindex_err_histogram} and \ref{fig:Mindex_offset_histogram}) and the median values of the activity index distributions (see Table~\ref{tab:median_of_measures}) suggest that the magnitudes of the activity indexes of different Balmer lines are at different levels.
To demonstrate the differences of the activity index magnitudes between the Balmer lines more clearly, 
in Fig.~\ref{fig:LMindex_with_teff}, the distributions of the activity indexes with stellar effective temperature ($T_\mathrm{eff}$) are shown for the stellar source sample of solar-like stars employed in this work.
The $T_\mathrm{eff}$ parameter used in Fig.~\ref{fig:LMindex_with_teff} is determined from the LRS data.   
The medians of the activity index distributions, as given in Table~\ref{tab:median_of_measures}, are indicated by the horizontal dashed lines in Fig.~\ref{fig:LMindex_with_teff}.

It can be seen from Fig.~\ref{fig:LMindex_with_teff} that, from the lower-order to higher-order Balmer lines,
the magnitudes of the activity indexes (represented by the medians of the index distributions; see the horizontal dashed lines in Fig.~\ref{fig:LMindex_with_teff}) first decrease from $L_\mathrm{H\alpha}$ index to $L_\mathrm{H\beta}$ index (see Figs.~\ref{fig:LMindex_with_teff}c and \ref{fig:LMindex_with_teff}d), 
and then gradually increase from $L_\mathrm{H\beta}$ index to $L_\mathrm{H\gamma}$ and $L_\mathrm{H\delta}$ indexes (see Figs.~\ref{fig:LMindex_with_teff}a--\ref{fig:LMindex_with_teff}c). 
The distributions of the $L_\mathrm{H\alpha}$ index (Fig.~\ref{fig:LMindex_with_teff}d) and the $M_\mathrm{H\alpha}$ index (Fig.~\ref{fig:LMindex_with_teff}e) are generally consistent,
indicating the consistency between the LRS and MRS data.
The much smaller uncertainty of the $M_\mathrm{H\alpha}$ data than that of the $L_\mathrm{H\alpha}$ data, as having been shown in Section \ref{sec:index}, are also demonstrated in Figs.~\ref{fig:LMindex_with_teff}d and \ref{fig:LMindex_with_teff}e.
Although the $T_\mathrm{eff}$ is limit to a narrow range for the solar-like sample employed in this work ($\pm 75$\,K around the solar value; see Section \ref{sec:data}),
the trend of the activity indexes of the four Balmer lines with $T_\mathrm{eff}$ can be seen in Fig.~\ref{fig:LMindex_with_teff},
that is, the bottom envelopes of the activity index distributions increase with the decrease of $T_\mathrm{eff}$ for all the four Balmer lines.
This trend is mainly caused by the variation of the continuum flux (the denominator in Equations (\ref{equ:lrs_Lindex}) and (\ref{equ:I_tilde_index})) with $T_\mathrm{eff}$,
and can be weakened by utilizing the absolute flux indexes defined in Section \ref{sec:absflux} (see Section \ref{sec:result_absindex} for the detailed analysis of the absolute flux index distributions).

\subsection{Distribution of the absolute flux indexes of Balmer lines} \label{sec:result_absindex}

\subsubsection{Distribution of the absolute flux indexes with stellar effective temperature} \label{sec:result_absindex_teff}

\begin{figure*}
  \centering
  \includegraphics[width=1.01\textwidth]{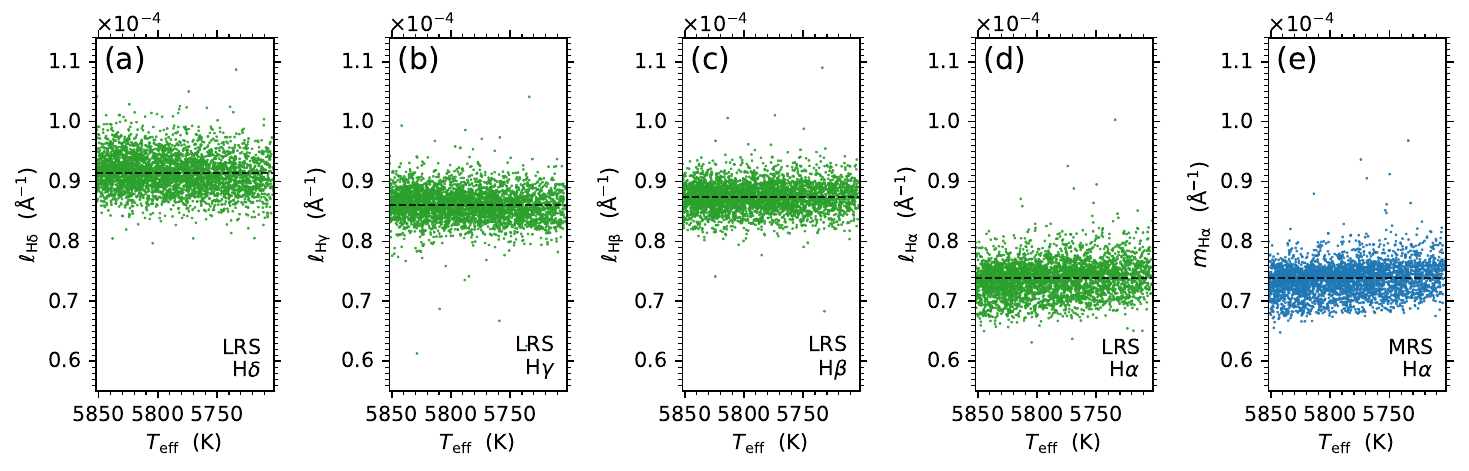}
  \caption{\scriptsize
  Distributions of the absolute flux indexes $\ell_\mathrm{H\alpha}$, $\ell_\mathrm{H\beta}$, $\ell_\mathrm{H\gamma}$, $\ell_\mathrm{H\delta}$ (from LRS data) and $m_\mathrm{H\alpha}$ (from MRS data) with $T_\mathrm{eff}$ for the stellar source sample of solar-like stars employed in this work.  
  The horizontal dashed line in each panel indicates the median of the absolute flux index distribution.
  The $T_\mathrm{eff}$ parameter used in the plots is determined from the LRS data.   
  }
  \label{fig:lmabsindex_with_teff}
\end{figure*}

The absolute flux indexes defined in Section \ref{sec:absflux} express the absolute fluxes at the centers of the Balmer lines with the ratios to the stellar bolometric flux, 
which counteracts the dependence of the activity indexes with $T_\mathrm{eff}$ shown in Fig.~\ref{fig:LMindex_with_teff},
and hence are more suitable for comparing activity properties of stars with different $T_\mathrm{eff}$ and intensity magnitudes of different Balmer lines.
The absolute activity indexes obtained in Section \ref{sec:absflux} include the $\ell_\mathrm{H\alpha}$, $\ell_\mathrm{H\beta}$, $\ell_\mathrm{H\gamma}$, and $\ell_\mathrm{H\delta}$ indexes from the LRS data and the $m_\mathrm{H\alpha}$ index from the MRS data.
The distributions of these absolute flux indexes with $T_\mathrm{eff}$ for the stellar source sample of solar-like stars employed in this work are shown in Fig.~\ref{fig:lmabsindex_with_teff}.
The $T_\mathrm{eff}$ parameter used in Fig.~\ref{fig:lmabsindex_with_teff} is determined from the LRS data.   
The medians of the absolute flux index distributions, as given in Table~\ref{tab:median_of_measures}, are indicated by the horizontal dashed lines in Fig.~\ref{fig:lmabsindex_with_teff}.

It can be seen from Fig.~\ref{fig:lmabsindex_with_teff} that the trend of the activity indexes with $T_\mathrm{eff}$ shown in Fig.~\ref{fig:LMindex_with_teff} has been largely weakened in the distribution plots of the absolute flux indexes with $T_\mathrm{eff}$. 
The magnitudes of the $\ell_\mathrm{H\beta}$, $\ell_\mathrm{H\gamma}$, and $\ell_\mathrm{H\delta}$ indexes (represented by the medians of the index distributions; see the horizontal dashed lines in Fig.~\ref{fig:lmabsindex_with_teff}) are at a similar level with small fluctuations among the lines (see Figs.~\ref{fig:lmabsindex_with_teff}a--\ref{fig:lmabsindex_with_teff}c), 
while the magnitude of the $\ell_\mathrm{H\alpha}$ index (see Fig.~\ref{fig:lmabsindex_with_teff}d) is lower than that of the other three lines.
The consistency between the $\ell_\mathrm{H\alpha}$ index (obtained from the LRS data) and the $m_\mathrm{H\alpha}$ index (obtained from the MRS data),
as well as the smaller uncertainty of the $m_\mathrm{H\alpha}$ data than the $\ell_\mathrm{H\alpha}$ data (as having been shown in Section \ref{sec:absflux}),
is demonstrated in Figs.~\ref{fig:lmabsindex_with_teff}d and \ref{fig:lmabsindex_with_teff}e.

\subsubsection{Distribution of the absolute flux indexes with stellar surface gravity} \label{sec:result_absindex_logg}

\begin{figure*}
  \centering
  \includegraphics[width=1.01\textwidth]{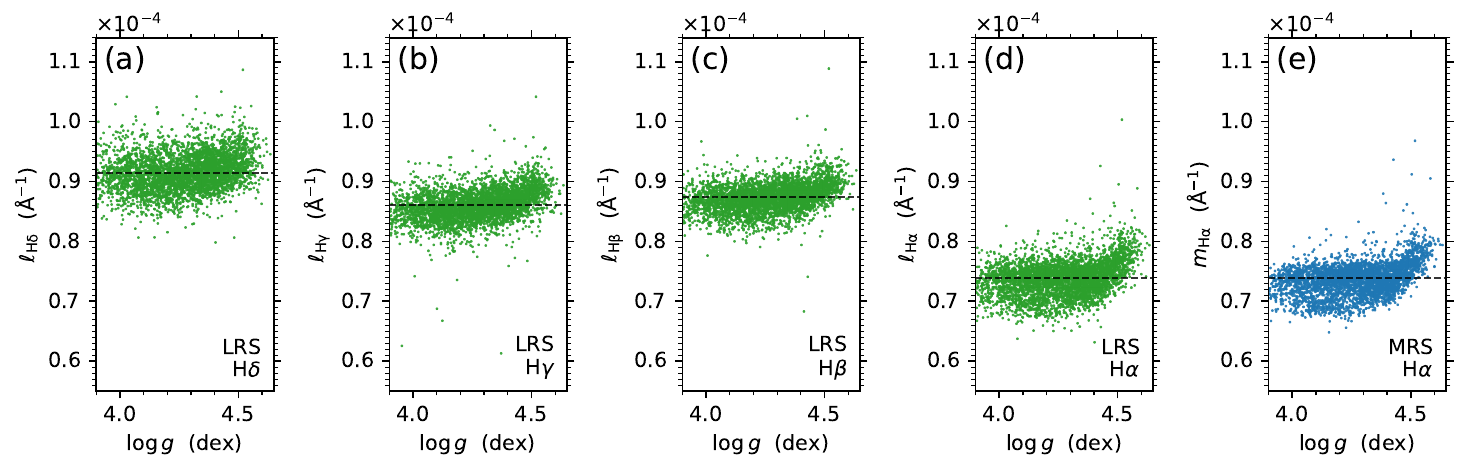}
  \caption{\scriptsize
  Distributions of the absolute flux indexes $\ell_\mathrm{H\alpha}$, $\ell_\mathrm{H\beta}$, $\ell_\mathrm{H\gamma}$, $\ell_\mathrm{H\delta}$ (from LRS data) and $m_\mathrm{H\alpha}$ (from MRS data) with stellar surface gravity ($\log g$) for the stellar source sample of solar-like stars employed in this work.  
  The horizontal dashed line in each panel indicates the median of the absolute flux index distribution.
  The $\log g$ parameter used in the plots is determined from the LRS data.   
  }
  \label{fig:lmabsindex_with_logg}
\end{figure*}

To investigate the dependence of the absolute flux indexes of the Balmer lines on stellar surface gravity ($\log g$),
in Fig.~\ref{fig:lmabsindex_with_logg}, the distributions of the absolute flux indexes $\ell_\mathrm{H\alpha}$, $\ell_\mathrm{H\beta}$, $\ell_\mathrm{H\gamma}$, $\ell_\mathrm{H\delta}$, and $m_\mathrm{H\alpha}$ with $\log g$ are shown for the stellar source sample of solar-like stars employed in this work.
The $\log g$ parameter used in Fig.~\ref{fig:lmabsindex_with_logg} is determined from the LRS data.   

It can be see from Figs.~\ref{fig:lmabsindex_with_logg}d and \ref{fig:lmabsindex_with_logg}e that,
for the H$\alpha$ line, 
there is a trend of the $\ell_\mathrm{H\alpha}$ or $m_\mathrm{H\alpha}$ index with the stellar surface gravity;
that is, the bottom envelope of the $\ell_\mathrm{H\alpha}$ or $m_\mathrm{H\alpha}$ distribution tends to be higher with the increase of stellar surface gravity for $\log g$ values greater than about $4.2$ dex, 
and there is an abrupt increase of $\ell_\mathrm{H\alpha}$ or $m_\mathrm{H\alpha}$ at $\log g$ values around about $4.5$ dex.
This trend can also be seen in the distribution plots of the absolute flux indexes of the other three Balmer lines ($\ell_\mathrm{H\beta}$, $\ell_\mathrm{H\gamma}$, and $\ell_\mathrm{H\delta}$; see Figs.~\ref{fig:lmabsindex_with_logg}a--\ref{fig:lmabsindex_with_logg}c),
but is not as clear as that of the H$\alpha$ line.

\subsubsection{Distribution of the absolute flux indexes with stellar metallicity} \label{sec:result_absindex_feh}

\begin{figure*}
  \centering
  \includegraphics[width=1.01\textwidth]{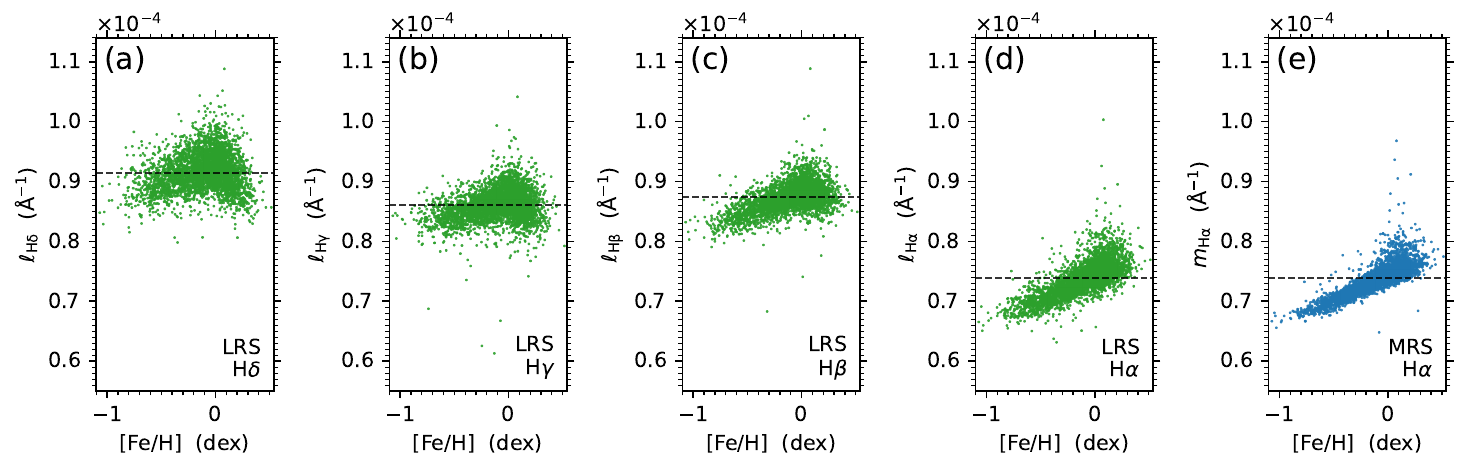}
  \caption{\scriptsize
    Distributions of the absolute flux indexes $\ell_\mathrm{H\alpha}$, $\ell_\mathrm{H\beta}$, $\ell_\mathrm{H\gamma}$, $\ell_\mathrm{H\delta}$ (from LRS data) and $m_\mathrm{H\alpha}$ (from MRS data) with stellar metallicity ($\mathrm{[Fe/H]}$) for the stellar source sample of solar-like stars employed in this work.  
  The horizontal dashed line in each panel indicates the median of the absolute flux index distribution.
  The $\mathrm{[Fe/H]}$ parameter used in the plots is determined from the LRS data.   
  }
  \label{fig:lmabsindex_with_feh}
\end{figure*}

The distributions of the $\ell_\mathrm{H\alpha}$, $\ell_\mathrm{H\beta}$, $\ell_\mathrm{H\gamma}$, $\ell_\mathrm{H\delta}$, and $m_\mathrm{H\alpha}$ indexes with stellar metallicity ($\mathrm{[Fe/H]}$) are displayed in Fig.~\ref{fig:lmabsindex_with_feh} to show the dependence of the absolute flux indexes of the Balmer lines on metallicity of stars. 
The $\mathrm{[Fe/H]}$ parameter used in Fig.~\ref{fig:lmabsindex_with_feh} is determined from the LRS data.   

It can be seen from Figs.~\ref{fig:lmabsindex_with_feh}d and \ref{fig:lmabsindex_with_feh}e that, for the H$\alpha$ line, 
there is a positive correlation trend between the $\ell_\mathrm{H\alpha}$ or $m_\mathrm{H\alpha}$ index and the stellar metallicity for the sample with lower $\ell_\mathrm{H\alpha}$ or $m_\mathrm{H\alpha}$ values (less than about $0.77 \times 10^{-4}$~{\AA}$^{-1}$),
and the sample with higher $\ell_\mathrm{H\alpha}$ or $m_\mathrm{H\alpha}$ values (greater than about $0.77 \times 10^{-4}$~{\AA}$^{-1}$) distributes around $\mathrm{[Fe/H]} \sim 0.1$~dex and forms a spike-shaped structure.
For the other three Balmer lines (see Figs.~\ref{fig:lmabsindex_with_feh}a--\ref{fig:lmabsindex_with_feh}c),
the trends of the relations between the absolute flux indexes ($\ell_\mathrm{H\beta}$, $\ell_\mathrm{H\gamma}$, and $\ell_\mathrm{H\delta}$) and the stellar metallicity are not monotonic,
and the shapes of the trends of the three Balmer lines are different from each other as exhibited in Figs.~\ref{fig:lmabsindex_with_feh}a--\ref{fig:lmabsindex_with_feh}c. 

\subsubsection{Relationship between the absolute flux indexes of Balmer lines} \label{sec:result_absindex_relation}

\begin{figure*}
  \centering
  \includegraphics[width=0.95\textwidth]{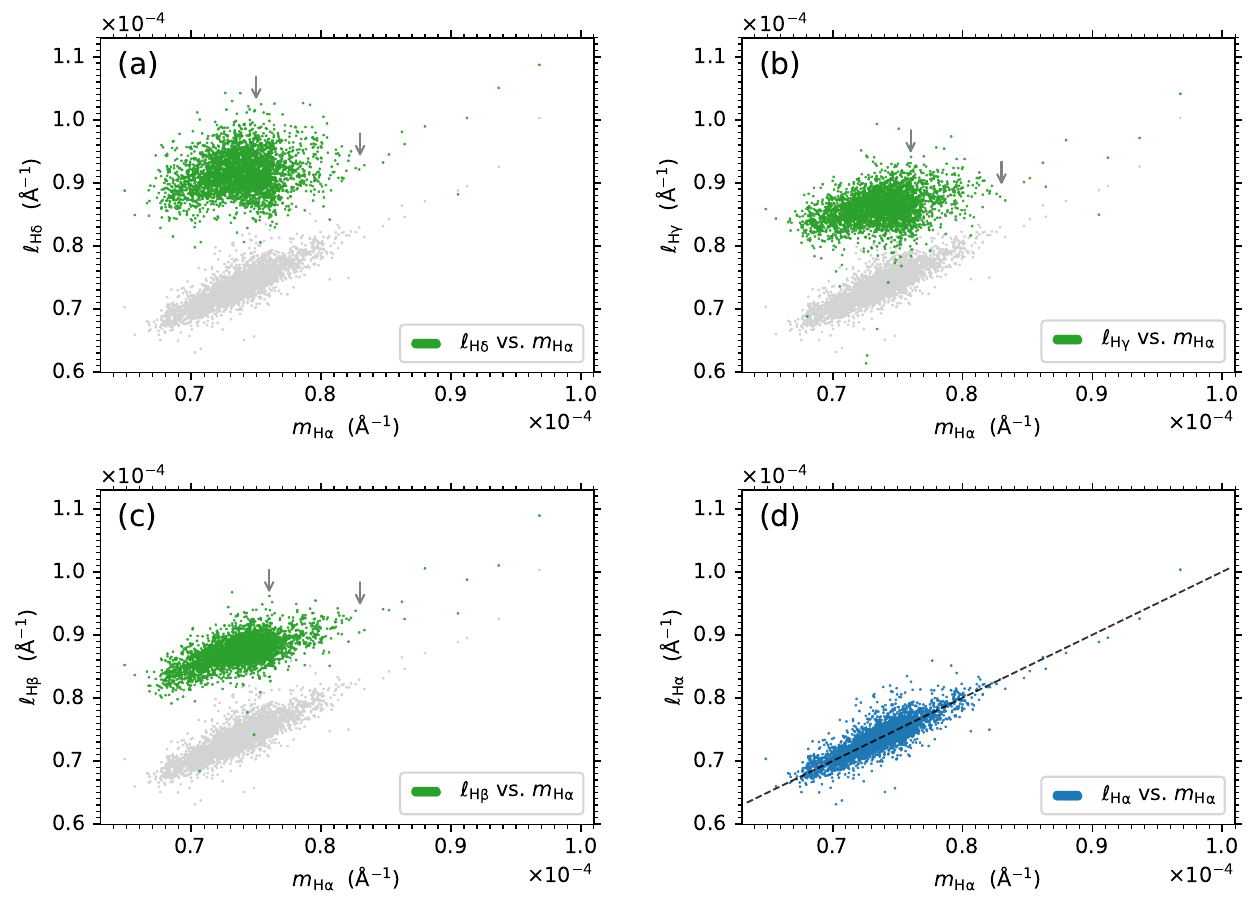}
  \caption{\scriptsize
  Scatter plots of the absolute flux indexes $\ell_\mathrm{H\alpha}$, $\ell_\mathrm{H\beta}$, $\ell_\mathrm{H\gamma}$, and $\ell_\mathrm{H\delta}$ (from LRS data) vs. $m_\mathrm{H\alpha}$ (from MRS data) for the stellar source sample of solar-like stars employed in this work.
  The scatter plot of $\ell_\mathrm{H\alpha}$ vs. $m_\mathrm{H\alpha}$ in panel (d) is also shown in panels (a)--(c) in gray for comparison.
  The $1:1$ line (dashed line) is displayed in panel (d) for reference.
  The short arrows in panels (a)--(c) indicate the turning positions of the upper envelopes of the data point distributions (see main text for details).
  }
  \label{fig:lmabsindex_relation}
\end{figure*}

The distributions of the absolute flux indexes of the four Balmer lines with stellar effective temperature, surface gravity, and metallicity described above show that the behavior of the H$\alpha$ line in response to stellar activity is very different from that of the H$\beta$, H$\gamma$, and H$\delta$ lines.  
In order to find out the relationship between the absolute flux indexes of the four Balmer lines,
in Fig.~\ref{fig:lmabsindex_relation}, the scatter plots of $\ell_\mathrm{H\alpha}$, $\ell_\mathrm{H\beta}$, $\ell_\mathrm{H\gamma}$, and $\ell_\mathrm{H\delta}$ indexes vs. $m_\mathrm{H\alpha}$ index are shown for the stellar source sample of solar-like stars employed in this work.
Because the uncertainty of the $m_\mathrm{H\alpha}$ index from MRS data is smaller than that of the $\ell_\mathrm{H\alpha}$ index from LRS data (see Section \ref{sec:absflux} and Table~\ref{tab:median_of_measures}),
the $m_\mathrm{H\alpha}$ index is used as the representative index of the H$\alpha$ line (see the horizontal axis of the plots in Fig.~\ref{fig:lmabsindex_relation}),
and its relations with the $\ell_\mathrm{H\beta}$, $\ell_\mathrm{H\gamma}$, and $\ell_\mathrm{H\delta}$ indexes of the other three Balmer lines are investigated (see Figs.~\ref{fig:lmabsindex_relation}a--\ref{fig:lmabsindex_relation}c).
The scatter plot of $\ell_\mathrm{H\alpha}$ vs. $m_\mathrm{H\alpha}$ is shown in Fig.~\ref{fig:lmabsindex_relation}d (as well as in other panels of Fig.~\ref{fig:lmabsindex_relation} in gray) for comparison. 

It can be seen from Fig.~\ref{fig:lmabsindex_relation}d that the data points of $\ell_\mathrm{H\alpha}$ vs. $m_\mathrm{H\alpha}$ are distributed mainly along the $1:1$ line (with a correlation coefficient of 0.86),
demonstrating the consistency between the $\ell_\mathrm{H\alpha}$ index (from LRS data) and the $m_\mathrm{H\alpha}$ index (from MRS data).
The scatter plots of $\ell_\mathrm{H\beta}$, $\ell_\mathrm{H\gamma}$, and $\ell_\mathrm{H\delta}$ vs. $m_\mathrm{H\alpha}$ in Figs.~\ref{fig:lmabsindex_relation}a--\ref{fig:lmabsindex_relation}c show that the relations between the absolute flux indexes of the $H\beta$, $H\gamma$, and $H\delta$ lines and the H$\alpha$ line are not monotonic.  
With the increase of the $m_\mathrm{H\alpha}$ index, 
the $\ell_\mathrm{H\beta}$, $\ell_\mathrm{H\gamma}$, and $\ell_\mathrm{H\delta}$ indexes first present trend of increasing and then deceasing, 
and finally increase synchronously with the $m_\mathrm{H\alpha}$ index
(see the short arrows in Figs.~\ref{fig:lmabsindex_relation}a--\ref{fig:lmabsindex_relation}c for the turning positions of the upper envelopes of the data point distributions).
The deviation from the positive correlation between the $\ell_\mathrm{H\beta}$, $\ell_\mathrm{H\gamma}$, and $\ell_\mathrm{H\delta}$ indexes and the $m_\mathrm{H\alpha}$ index becomes greater from the lower-order to higher-order Balmer lines (with the correlation coefficients of $\ell_\mathrm{H\beta}$, $\ell_\mathrm{H\gamma}$, and $\ell_\mathrm{H\delta}$ vs. $m_\mathrm{H\alpha}$ being 0.65, 0.37, and 0.22, respectively).
The different behaviors of the Balmer lines in response to stellar activity imply their different forming mechanisms.
This issue will be discussed in Section \ref{sec:discuss_balmer_lines}.

\section{Discussion} \label{sec:discuss}

\subsection{Balmer lines for stellar chromospheric activity} \label{sec:discuss_balmer_lines}

The analysis results in Section \ref{sec:results} reveal the different behavior of the Balmer H$\alpha$ line from that of the H$\beta$, H$\gamma$, and H$\delta$ lines in response to stellar activity:

\begin{enumerate}

\item The magnitudes of the activity indexes $L_\mathrm{H\beta}$, $L_\mathrm{H\gamma}$, and $L_\mathrm{H\delta}$ constitute an ascending sequence (see Fig.~\ref{fig:LMindex_with_teff}),
and the magnitude of the activity index of the H$\alpha$ line ($L_\mathrm{H\alpha}$ or $M_\mathrm{H\alpha}$) is not in this sequence.

\item The magnitudes of the absolute flux indexes $\ell_\mathrm{H\beta}$, $\ell_\mathrm{H\gamma}$, and $\ell_\mathrm{H\delta}$ are at the similar level (see Fig.~\ref{fig:lmabsindex_with_teff}),
which is well above the magnitude of the absolute flux index of the H$\alpha$ line ($\ell_\mathrm{H\alpha}$ or $m_\mathrm{H\alpha}$).

\item The distribution of the $\ell_\mathrm{H\alpha}$ or $m_\mathrm{H\alpha}$ index with stellar surface gravity has an abrupt increase at $\log g$ values around about $4.5$ dex (see Fig.~\ref{fig:lmabsindex_with_logg}),
while in the distribution plots of the $\ell_\mathrm{H\beta}$, $\ell_\mathrm{H\gamma}$, and $\ell_\mathrm{H\delta}$ indexes with $\log g$, this trend is not as clear as in the distribution plot of the H$\alpha$ line.

\item In the distribution plot of the $\ell_\mathrm{H\alpha}$ or $m_\mathrm{H\alpha}$ index with stellar metallicity, 
the H$\alpha$ absolute flux index has a positive correlation trend with $\mathrm{[Fe/H]}$ for the sample with lower index values,
and the distribution presents a spike-shaped structure for the sample with higher index values (see Fig.~\ref{fig:lmabsindex_with_feh}),
while in the distribution plots of the $\ell_\mathrm{H\beta}$, $\ell_\mathrm{H\gamma}$, and $\ell_\mathrm{H\delta}$ indexes with $\mathrm{[Fe/H]}$, the trends of the indexes with $\mathrm{[Fe/H]}$ are not monotonic.

\item The relations between the $\ell_\mathrm{H\beta}$, $\ell_\mathrm{H\gamma}$, and $\ell_\mathrm{H\delta}$ indexes and the H$\alpha$ absolute flux index are not monotonic (see Fig.~\ref{fig:lmabsindex_relation}), 
and the deviation from the positive correlation becomes greater for higher-order Balmer lines.

\end{enumerate}

The different behaviors of the H$\alpha$ line compared with those of the H$\beta$, H$\gamma$, and H$\delta$ lines described above suggest that the forming mechanism of the H$\alpha$ line-core intensity is different from that of the other three Balmer lines.
According to the modeling results by Vernazza et al., Fontenla et al., and Avrett et al. \citep{1973ApJ...184..605V, 1981ApJS...45..635V, 1999ApJ...518..480F, 2008ApJS..175..229A} (see introduction in Section~\ref{sec:intro}),
the line-core intensity of the H$\alpha$ line of solar-like stars is formed in the chromosphere,
while the central intensities of the H$\beta$ and H$\gamma$ lines have substantial contributions from the photosphere;
thus, it is natural that the line-core intensity of the H$\alpha$ line is a good indicator of the chromospheric activity, 
and the central intensities of the H$\beta$ and H$\gamma$ lines can not well represent the chromospheric activity.
The behavior of the H$\delta$ line deviates from the H$\alpha$ line even larger than the H$\beta$ and H$\gamma$ lines, 
implying that the photospheric contribution to the central intensity of the H$\delta$ line is even greater than that of the H$\beta$ and H$\gamma$ lines.

According to the modeling results by Cram and Mullan \citep{1979ApJ...234..579C, 1985ApJ...294..626C} for spectral profiles of the Balmer lines of cool stars (see introduction in Section \ref{sec:intro}),
when the chromosphere is absent, 
the Balmer lines manifest as weak absorption lines (corresponding to a larger value of absolute flux index); 
as the chromosphere materials increase, the Balmer lines first become deeper (decrease of index value), then fill in to become shallower (increase of index value).
Since the line-core intensity of the H$\alpha$ line is formed in the chromosphere, 
the rule of first decrease and then increase of line-core intensity does not apply to the H$\alpha$ line.
However, the phenomena of decrease of central intensity with increase of stellar activity do exist for the H$\beta$, H$\gamma$, and H$\delta$ lines as revealed by the results in Section \ref{sec:result_absindex} 
(see the trend between the two short arrows in Figs.~\ref{fig:lmabsindex_relation}a--\ref{fig:lmabsindex_relation}c),
which imply that, with the increase of stellar activity, 
more and more chromospheric materials participate in the forming of the central intensities of the H$\beta$, H$\gamma$, and H$\delta$ lines (in conjunction with the photospheric contribution), resulting in a deeper line profile, 
and finally the chromosphere dominates the central intensities of the lines,
leading to a synchronous variation with the line-core intensity of the H$\alpha$ line (see the trend on the right of the second arrow in Figs.~\ref{fig:lmabsindex_relation}a--\ref{fig:lmabsindex_relation}c).

From the above discussion, it can be seen that the best indicator of the chromospheric activity of solar-like stars in the Balmer series is the H$\alpha$ line.
The responses of the central intensities of the H$\beta$, H$\gamma$, and H$\delta$ lines to stellar activity are complex and the trends are not monotonic;
nevertheless, the central intensities of the lines can help to reveal the stratification of the activity levels of solar-like stars as discussed in Section \ref{sec:discuss_activity_stratification}.

\subsection{Stratification of the activity levels of solar-like stars} \label{sec:discuss_activity_stratification}

\begin{figure*}
  \centering
  \includegraphics[width=1.01\textwidth]{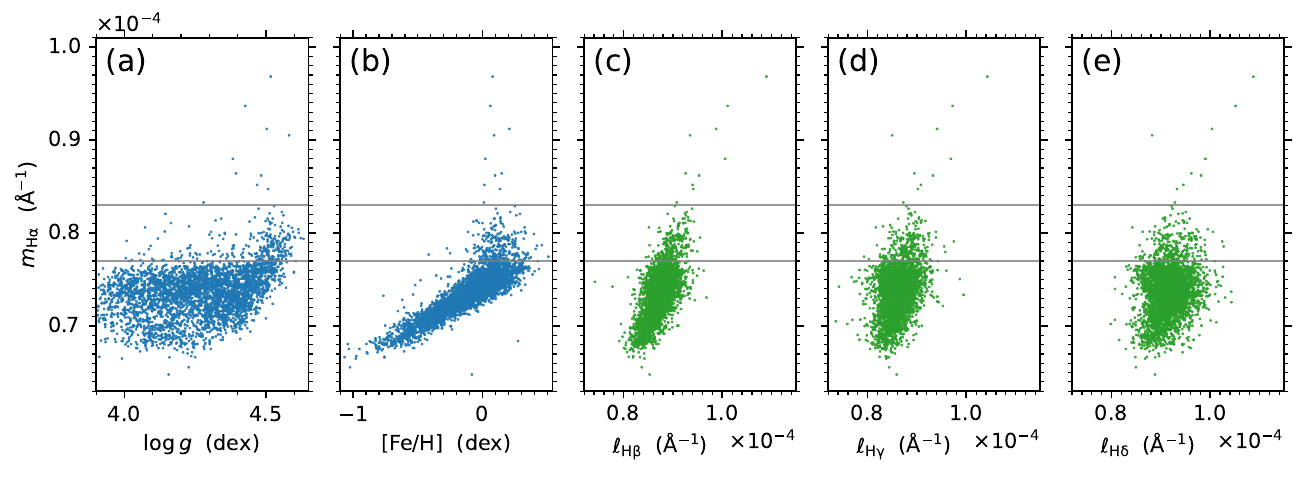}
  \caption{\scriptsize
  Diagram illustrating the stratification of the activity levels of solar-like stars revealed by the distributions of the absolute flux indexes of the Balmer lines.
  (a--e) Scatter plots of $m_\mathrm{H\alpha}$ vs. $\log g$, $\mathrm{[Fe/H]}$, $\ell_\mathrm{H\beta}$, $\ell_\mathrm{H\gamma}$, and $\ell_\mathrm{H\delta}$, respectively.
  The two horizontal lines represent $m_\mathrm{H\alpha} = 0.77 \times 10^{-4}$  {\AA}$^{-1}$ and $0.83 \times 10^{-4}$ {\AA}$^{-1}$,
  which roughly divide the sample of solar-like stars into three distinct activity stages (see main text for details).
  }
  \label{fig:absindex_stratification}
\end{figure*}

In addition to the different behaviors of the Balmer lines in response to stellar activity discussed in Section \ref{sec:discuss_balmer_lines},
the distributions of the absolute flux indexes of the Balmer lines presented in Section \ref{sec:result_absindex} also reveal the stratification of the activity levels of solar-like stars,
which is illustrated in Fig.~\ref{fig:absindex_stratification}.
The five panels of Fig.~\ref{fig:absindex_stratification} reproduce the distributions of $m_\mathrm{H\alpha}$ vs. $\log g$, $\mathrm{[Fe/H]}$, $\ell_\mathrm{H\beta}$, $\ell_\mathrm{H\gamma}$, and $\ell_\mathrm{H\delta}$ obtained in Section \ref{sec:result_absindex}.
(Since the $m_\mathrm{H\alpha}$ and $\ell_\mathrm{H\alpha}$ indexes are quantitatively consistent with each other,
the $m_\mathrm{H\alpha}$ index is utilized in the following discussion; 
the results are equally applicable to the $\ell_\mathrm{H\alpha}$ index.)
The stratification of the activity levels is most evident in the distribution plot of $m_\mathrm{H\alpha}$ vs. $\log g$ (Fig.~\ref{fig:absindex_stratification}a).
The two horizontal lines in the plot (corresponding to $m_\mathrm{H\alpha} = 0.77 \times 10^{-4}$ {\AA}$^{-1}$ and $0.83 \times 10^{-4}$ {\AA}$^{-1}$, respectively; also shown in other panels of Fig.~\ref{fig:absindex_stratification}) roughly divide the sample of solar-like stars into three distinct activity stages,
which are named normal stage, intense stage, and extremely intense stage, respectively, from bottom to top.

The normal activity stage of solar-like stars is featured by the flat upper envelope in the distribution plot of $m_\mathrm{H\alpha}$ vs. $\log g$ (see Fig.~\ref{fig:absindex_stratification}a) and the positive correlation trend between $m_\mathrm{H\alpha}$ and $\mathrm{[Fe/H]}$ (see Fig.~\ref{fig:absindex_stratification}b);
the relations of $m_\mathrm{H\alpha}$ vs. $\ell_\mathrm{H\beta}$, $\ell_\mathrm{H\gamma}$, and $\ell_\mathrm{H\delta}$ are not monotonic in the normal activity stage as shown in Figs.~\ref{fig:absindex_stratification}c--\ref{fig:absindex_stratification}e.
Most stars in the solar-like sample are in the normal stage as shown in Fig.~\ref{fig:absindex_stratification} (with the proportion of about 92.0\%). 
The intense stage is featured by the abrupt increase of $m_\mathrm{H\alpha}$ at $\log g$ values around about $4.5$ dex (see Fig.~\ref{fig:absindex_stratification}a) and the spike-shaped structure in the $m_\mathrm{H\alpha}$ vs. $\mathrm{[Fe/H]}$ plot around $\mathrm{[Fe/H]} \sim 0.1$~dex (see Fig.~\ref{fig:absindex_stratification}b). 
The number of stars in the intense stage (about 7.7\%) is less than that in the normal stage. 
The extremely intense stage is featured by the much higher activity level than the other two stages, 
and only very sparse data points are in the extremely intense stage (about 0.3\%).
The distinction between the extremely intense stage and the intense stage is most evident in the plots of $m_\mathrm{H\alpha}$ vs. $\ell_\mathrm{H\beta}$, $\ell_\mathrm{H\gamma}$, and $\ell_\mathrm{H\delta}$ (Figs.~\ref{fig:absindex_stratification}c--\ref{fig:absindex_stratification}e),
that is, the dividing line of $m_\mathrm{H\alpha} = 0.83 \times 10^{-4}$~{\AA}$^{-1}$ just corresponds to the second turning position of the distributions (see the second short arrow in Figs.~\ref{fig:lmabsindex_relation}a--\ref{fig:lmabsindex_relation}c), 
above which $\ell_\mathrm{H\beta}$, $\ell_\mathrm{H\gamma}$, and $\ell_\mathrm{H\delta}$ increase synchronously with $m_\mathrm{H\alpha}$.
The stratification of the activity levels of solar-like stars implies the very different magnetic field environments and physical conditions associated with the different stages of stellar activity.

\section{Summary and Conclusion} \label{sec:conclusion}

In this paper, the intensities of the Balmer lines are investigated for stellar chromospheric activity of solar-like stars by using the LRS and MRS spectral data of LAMOST.
The Balmer lines analyzed are the H$\alpha$, H$\beta$, H$\gamma$, and H$\delta$ lines in the LRS data and the H$\alpha$ line in the MRS data.
The high-S/N and co-source spectra of LRS and MRS are used in the analysis.
The span of the stellar effective temperature of the solar-like spectral sample employed in this work is $\pm 75$\,K around the solar value;
the stellar surface gravity of the solar-like sample is $\ge 3.9$~dex;
and no further restriction on stellar metallicity is imposed.

The activity indexes $L_\mathrm{H\alpha}$, $L_\mathrm{H\beta}$, $L_\mathrm{H\gamma}$, and $L_\mathrm{H\delta}$ of the four Balmer lines in the LRS data are introduced and evaluated for the LRS solar-like spectral sample employed in this work,
which are defined as the ratio of the mean flux in the central band of the Balmer lines to the weighted mean flux in the two continuum bands on the two sides of the lines;
the activity index $M_\mathrm{H\alpha}$ is also defined and evaluated based on the smoothed MRS data for the MRS solar-like spectral sample employed in this work.
The value of the $M_\mathrm{H\alpha}$ index is calibrated with the value of the $L_\mathrm{H\alpha}$ index to acquire quantitative consistency between the two activity indexes obtained from the MRS and LRS data.
Relative to the $L_\mathrm{H\alpha}$ index, the advantage of the $M_\mathrm{H\alpha}$ index is that its uncertainty is about one order of magnitude smaller than the uncertainty of the $L_\mathrm{H\alpha}$ index.
The distributions of the activity indexes of the Balmer lines show that the magnitudes of the $L_\mathrm{H\beta}$, $L_\mathrm{H\gamma}$, and $L_\mathrm{H\delta}$ indexes (represented by the medians of the index distributions) constitute an ascending sequence (i.e., $L_\mathrm{H\beta} < L_\mathrm{H\gamma} < L_\mathrm{H\delta}$), 
and the magnitude of the H$\alpha$ activity index ($L_\mathrm{H\alpha}$ or $M_\mathrm{H\alpha}$) is not in this sequence.

On the basis of the activity indexes of the Balmer lines,
the absolute fluxes at the centers of the Balmer lines are further evaluated, 
and the absolute flux indexes $\ell_\mathrm{H\alpha}$, $\ell_\mathrm{H\beta}$, $\ell_\mathrm{H\gamma}$, and $\ell_\mathrm{H\delta}$ for the LRS data and $m_\mathrm{H\alpha}$ for the MRS data are introduced, 
which are defined as the ratio of the Balmer line absolute flux to the stellar bolometric flux.
The distributions of the absolute flux indexes of the Balmer lines show that the magnitudes of the $\ell_\mathrm{H\beta}$, $\ell_\mathrm{H\gamma}$, and $\ell_\mathrm{H\delta}$ indexes are at the similar level,
and the magnitude of the H$\alpha$ absolute flux index ($\ell_\mathrm{H\alpha}$ or $m_\mathrm{H\alpha}$) is lower than that of other three Balmer lines.
In the distributions of the absolute flux indexes with $\log g$,
the $\ell_\mathrm{H\alpha}$ or $m_\mathrm{H\alpha}$ index has an abrupt increase at $\log g$ values around about $4.5$ dex,
and this feature is not clear in the distribution plots the $\ell_\mathrm{H\beta}$, $\ell_\mathrm{H\gamma}$, and $\ell_\mathrm{H\delta}$ indexes.
In the distributions of the absolute flux indexes with $\mathrm{[Fe/H]}$,
the $\ell_\mathrm{H\alpha}$ or $m_\mathrm{H\alpha}$ index has a positive correlation trend with $\mathrm{[Fe/H]}$ for the sample with lower index values,
and the distribution presents a spike-shaped structure for the sample with higher index values;
whereas the trends between the $\ell_\mathrm{H\beta}$, $\ell_\mathrm{H\gamma}$, and $\ell_\mathrm{H\delta}$ indexes and $\mathrm{[Fe/H]}$ are not monotonic.
The correlation analysis between the $\ell_\mathrm{H\beta}$, $\ell_\mathrm{H\gamma}$, and $\ell_\mathrm{H\delta}$ indexes and the H$\alpha$ absolute flux index (represented by the $m_\mathrm{H\alpha}$ index for its smaller uncertainty than the $\ell_\mathrm{H\alpha}$ index) shows that 
the relations between the $\ell_\mathrm{H\beta}$, $\ell_\mathrm{H\gamma}$, and $\ell_\mathrm{H\delta}$ indexes and the $m_\mathrm{H\alpha}$ index are not monotonic;
with the increase of the $m_\mathrm{H\alpha}$ index, the $\ell_\mathrm{H\beta}$, $\ell_\mathrm{H\gamma}$, and $\ell_\mathrm{H\delta}$ indexes first present trend of increasing and then decreasing, and finally increase synchronously with the $m_\mathrm{H\alpha}$ index,
and the deviation from the positive correlation is greater for higher-order Balmer lines.

The different behavior of the H$\alpha$ line from that of the H$\beta$, H$\gamma$, and H$\delta$ lines in response to stellar magnetic activity can be interpreted by the different mechanisms by which the line-core intensities of the Balmer lines are formed. 
By referring to the modeling results for solar atmosphere \citep{1973ApJ...184..605V, 1981ApJS...45..635V, 1999ApJ...518..480F, 2008ApJS..175..229A},
the line-core intensity of the H$\alpha$ line is formed in the chromosphere and hence is a good indicator of stellar chromospheric activity;
whereas the central intensities of the H$\beta$, H$\gamma$, and H$\delta$ lines have substantial contributions from the photosphere, 
as a result, the responses of the central intensities of the three Balmer lines to stellar activity are complex and not monotonic.

The distributions of the absolute flux indexes of the Balmer lines also reveal the stratification of the activity levels of solar-like stars.
The three distinct activity stages (normal stage, intense stage, and extremely intense stage) from lower to higher activity levels are roughly divided by $m_\mathrm{H\alpha}$ (or $\ell_\mathrm{H\alpha}$) $= 0.77 \times 10^{-4}$~{\AA}$^{-1}$ and $0.83 \times 10^{-4}$~{\AA}$^{-1}$. 
The normal stage includes the most stars in the solar-like sample;
it is featured by the flat upper envelope of the distribution of the $m_\mathrm{H\alpha}$ (or $\ell_\mathrm{H\alpha}$) index with $\log g$ and the positive correlation trend of $m_\mathrm{H\alpha}$ (or $\ell_\mathrm{H\alpha}$) with $\mathrm{[Fe/H]}$.
The intense stage is featured by the abrupt increase of $m_\mathrm{H\alpha}$ (or $\ell_\mathrm{H\alpha}$) at $\log g$ values around about $4.5$ dex and the spike-shaped structure around $\mathrm{[Fe/H]} \sim 0.1$~dex in the distribution of $m_\mathrm{H\alpha}$ (or $\ell_\mathrm{H\alpha}$) vs. $\mathrm{[Fe/H]}$.
The extremely intense stage is featured by the much higher activity level than the other two stages, 
and only very sparse data points in the solar-like sample are in this stage;
it is the activity stage in which the $\ell_\mathrm{H\beta}$, $\ell_\mathrm{H\gamma}$, and $\ell_\mathrm{H\delta}$ indexes increase synchronously with the $m_\mathrm{H\alpha}$ (or $\ell_\mathrm{H\alpha}$) index.
The distinct characteristics of the activity stages imply very different magnetic field environments and physical conditions of solar-like stars.

The approach developed in this work for the quantitative calibration between the activity measures of the LRS spectra and MRS spectra can be applied to further collaborative analyses of the LRS and MRS data for understanding stellar chromospheric activity.

\section*{Acknowledgements}
This research is supported by the National Key R\&D Program of China (2019YFA0405000) and the National Natural Science Foundation of China (11973059).
Guoshoujing Telescope (the Large Sky Area Multi-Object Fiber Spectroscopic Telescope, LAMOST) is a National Major Scientific Project built by the Chinese Academy of Sciences. Funding for the project has been provided by the National Development and Reform Commission. LAMOST is operated and managed by the National Astronomical Observatories, Chinese Academy of Sciences.
This work made use of Astropy \citep{2013A&A...558A..33A,2018AJ....156..123A} and SciPy \citep{2020NatMe..17..261V}.

\section*{Data availability} 
The data generated during the current study are available in the online dataset of this paper.

\end{document}